\documentclass[10pt,aps,prd,twocolumn,notitlepage,nofootinbib,floatfix,superscriptaddress,longbibliography]{revtex4-1}

\usepackage{amsmath, amsthm, amssymb, amsfonts, amsbsy, mathrsfs}

\usepackage{xcolor}
\usepackage{calligra,bm}
\usepackage{stmaryrd}
\usepackage{multirow}

\usepackage{graphicx}
\graphicspath{{./images/}}

\usepackage[hidelinks,bookmarks=true]{hyperref}
\hypersetup{pdfstartview=FitH,pdfhighlight=/O,colorlinks=false}

\begin{document}

\title{Relativistic location algorithm in curved spacetime}

\author{Justin C. Feng}
\email{justin.feng@tecnico.ulisboa.pt}
\affiliation{ CENTRA, Departamento de F{\'i}sica, Instituto Superior
              T{\'e}cnico – IST, Universidade de Lisboa – UL, Avenida
              Rovisco Pais 1, 1049 Lisboa, Portugal. }

\author{Filip Hejda}
\email{hejdaf@fzu.cz}
\affiliation{ CEICO, Institute of Physics of the Czech Academy of
        Sciences, Na Slovance 1999/2, 182 21 Prague 8, Czech Republic }

\author{Sante Carloni}
\email{sante.carloni@unige.it}
\affiliation{ DIME Sez. Metodi e Modelli Matematici, Universit\`{a} di
Genova,\\
              Via All'Opera Pia 15, Genoa, 16145, (Italy). }

%
%
\begin{abstract}
    In this article, we describe and numerically implement a method for
    relativistic location in slightly curved, but otherwise generic
    spacetimes. For terrestrial positioning in the context of Global
    Navigation Satellite Systems, our algorithm incorporates
    gravitational as well as tropospheric and ionospheric effects
    modeled by the Gordon metric. The algorithm is implemented in the
    \href{https://github.com/justincfeng/squirrel.jl}{\textsc{squirrel.jl}}
    code, which employs a quasi-Newton Broyden algorithm in conjunction
    with automatic differentiation of numerical geodesics. Our work
    provides a practical solution to the relativistic location problem
    in a generic spacetime and consolidates relativistic and atmospheric
    effects in a single framework. Though optimization is not our
    primary focus, our implementation is already fast enough for
    practical use, establishing a position from five emission points in
    $< 1~{\rm s}$ on a desktop computer for reasonably simple spacetime
    geometries. In vacuum, our implementation can achieve submillimeter
    accuracy considering the Kerr metric with terrestrial parameters and
    submeter accuracy including tropospheric and ionospheric effects.
\end{abstract}


\maketitle


%
%

%
%

\section{Introduction}\label{sec:I} Global Navigation Satellite Systems
(GNSSs) have become an indispensable tool in modern life. From civil
aviation to ridesharing, the applications of GNSSs continue to increase
in scope and usage. The increasing dependence of our modern economy on
GNSSs has led to the development of an expansive infrastructure aimed at
achieving more reliable, accurate, and precise location systems. 

Traditional GNSSs are based on a Newtonian framework, particularly on
the simple principle of trilateration in Euclidean space, i.e., the use
of three sources to determine the position of a given user. However, a
purely Newtonian framework is not enough; when one proceeds naively with
the calculation of the position of the user employing standard Newtonian
mechanics, even neglecting sources of errors associated with the signal
transmission, one is faced with large accumulative errors
\cite{Ashby2002,Ashby:2003vja}. Such errors are mainly sourced by two
effects. The first is the difference in clock rates due to the relative
motion of the user and the satellites, while the second is due to the
gravitational time dilation effects; the latter contribution is more
than six times larger than the former. Combined with other relativistic
effects, they amount to about a 40 microsecond delay per day. Translated
into location error, this offset would amount to an error of about 10 km
for every day of activity of the GNSS system. Correcting for
relativistic effects is therefore crucial for achieving an accurate
positioning system. At present, the relativistic offset is compensated
by simply designing the clocks on the satellites to be slower by about
40 microseconds (increasing the number of emitters aside). In addition,
the ground stations and receivers have to be provided with a
microcomputer able to process any additional calculation required and to
periodically reset the positioning system
\cite{Ashby2002,Ashby:2003vja}. Relativistic corrections therefore
increase the size of the ground GNSS infrastructure (see, e.g.,
\cite{teunissen2017springer,ISGPS200M} for some details in this matter),
which in turn increases the general cost and maintenance burden of the
system itself. 

In this context, it makes sense to design a positioning system based
directly on relativistic principles. The concept of a relativistic
positioning system (RPS) employs emission coordinates as the primary
coordinates for spacetime \cite{Coll:2006gv,Coll:2006sg}. Emission
coordinates are formed from the timestamps of proper time broadcasts for
a system of satellites, so that the location of the user in emission
coordinates is immediately established upon signal reception. Moreover,
the satellite coordinate positions become trivial in emission
coordinates, consisting of the satellite clock times and timestamps of
concurrently received signals. Of course, what is less trivial is the
transformation to a standard coordinate system and the specification of
satellite and user positions in terms of physical distances.

The simplicity of an RPS based on emission coordinates offers several
advantages over traditional implementations of GNSSs. Since the user and
satellite positions in emission coordinates are expressed directly in
terms of proper time broadcasts received by the users and satellites, an
implementation of an RPS in terms of emission coordinates has the
potential to reduce the post processing, number of emitters, and number
of ground stations, which would permit a significant reduction in the
size and scope of the infrastructure required without compromising (and
possibly improving) the performance and accuracy of the service.
Additionally, an RPS might be employed equally well for positioning in
space. Finally, RPSs can be used as key scientific tools; there are, for
instance, proposals to use an RPS network for relativistic geodesy as
well as the detection of gravitational waves (see, e.g.,
\cite{auge2011gravitational,*Delva:2011zk}). 

In recent years, efforts in the definition and development of a
consistent RPS has led to the development of a number of different
approaches
\cite{Pascual-Sanchez:1997zdx,Coll:2017way,Puchades:2016myk,Puchades:2013uzt,Puchades:2011bz,Coll:2010xy,Coll:2009zz,Pascual-Sanchez:2007lin,Coll:2006wk,Pozo:2006ps,MoralesLladosa:2006ux,Lachieze-Rey:2006lja,Colmenero:2021fbq,Ruggiero:2022,Fidalgo2021,Bahder:2003gm,Bahder:2001rz,Blagojevic:2001gt,Kostic:2015cca,Carloni:2018cuf,Ruggiero:2010nd,Tartaglia:2010sw,Tartaglia:2012rp}.
Much effort has been devoted to establishing a transformation between
emission coordinates and a standard coordinate system. The majority of
the approaches in this direction are limited to a small class of
geometrical backgrounds and require the inversion of transcendental
equations. One exception is that of \cite{Bunandar:2011kp}, which is
applicable for general backgrounds, but this approach still requires
numerically solving the (curved spacetime) Eikonal equation, a partial
differential equation. Thus a key point in the development of RPSs
is the development of calculational methods applicable to more general
spacetime geometries that are efficient enough to be performed on
standard hardware such as that available in handheld devices or
satellites. 

Another issue that is often neglected in the development of RPSs is the
modeling of nongravitational effects, such as the interaction of the
signal with the troposphere and ionosphere. These phenomena are
typically thought to require methods independent of the general
relativistic formalism. For this reason, despite the relevance of the
phenomena to the performance of the positioning system, and the fact
that they are among the largest contributors to typical GNSS error
budgets \cite{EU2016} (see for instance Tables 24 and 25 therein for 
typical error budgets), they are often excluded in the framework of 
RPSs. 

In this paper, we propose a new approach to the relativistic location
problem, applicable in generic, slightly curved, spacetimes, which can
by way of analog gravity models incorporate the interaction of light
signals with the troposphere and ionosphere in a fully relativistic
framework. We will then compare the performance of our method with
respect to the standard performance of the Galileo system, showing that
our method can in principle achieve similar results. Our approach
requires solving at minimum four ordinary differential equations (ODEs),
greatly reducing the computational complexity of calculations compared
to partial-differential-equation-based approaches. Our work can be seen
as complementary to the recent work \cite{Colmenero:2021fbq} and earlier
works \cite{Delva:2010zx,Cadez2010,Gomboc2014,Kostic:2015cca} that
address instead satellite ephemeris errors (another large contributor to
GNSS error budgets), which we neglect here. 

In the following, lists of symbols contained in the curly brackets
$\{x_1,x_2,...\}$ denote sets, and lists of more than two symbols
contained in the round brackets $(v_1,v_2,...)$ denote vectors, with
$v_I$ either representing components or lower-dimensional vectors. In
the latter case, $(v_1,v_2,...)$ represents a vector formed from the
concatenation of vectors $v_1,v_2,$ etc. Greek indices represent
spacetime coordinate indices and take values from the set $\{0,1,2,3\}$.
Lowercase latin indices from the middle of the alphabet $\{i,j,k,l\}$
represent spatial coordinate indices and take values from the set
$\{1,2,3\}$. Unless otherwise indicated, Einstein summation convention
is employed on coordinate indices. Uppercase latin indices ($I$, for
instance) and the lowercase latin indices $\{a,b\}$ are not
treated as tensor indices, and are used to label emitters and emission
points; the uppercase indices take values from the set $\{1,2,...,N\}$,
and the lowercase indices $\{a,b\}$ take values from the set
$\{1,2,3\}$. Lowercase bold latin letters (such as $\bf b$, $\bf v$, and
$\bf x$) are reserved for three-component quantities; when components of
such letters are displayed explicitly (for instance ${\bf x}^1$, ${\bf
x}^2_3$), raised indices always represent the value of the coordinate
index and the lowered indices represent the value of the emission point
label. Uppercase bold latin letters (such as $\bf A$ and $\bf J$) are
reserved for matrices.

In Sec. \ref{sec:II}, we discuss the problem of relativistic location in
flat spacetime. Our algorithm for relativistic location in curved
spacetime is described in Sec. \ref{sec:III}. In Secs. \ref{sec:IV} and
\ref{sec:V}, we describe the spacetime metrics and index of refraction
models used in tests of our implementation of the algorithm. Tests and
benchmarks of our implementation are described in Sec. \ref{sec:VI}. We
conclude with a summary and brief discussion in Sec. \ref{sec:VII}. 


%
%

%
%
\section{Relativistic location in flat spacetime} \label{sec:II}

%
%
\subsection{Relativistic positioning and relativistic location}
Relativistic positioning systems are based on the concept of emission
coordinates (a detailed discussion of which may be found in
\cite{Coll:2006gv,Coll:2006sg}; see also \cite{Coll:2006nz} for the
two-dimensional case), which correspond to the broadcasted proper times
of a system of at least four satellites. Each value of proper time
$\tau_I$ broadcasted by a satellite $I$ defines a (null) hypersurface
corresponding to events at which an observer receives the broadcasted
value $\tau_I$; this surface forms the future pointing light cone for
the spacetime position $X^\mu_I$ of satellite $I$ at the moment the
broadcast is emitted. Given four satellites, each with a single
broadcast of proper time (which we collectively write as
$\underline{\tau}=\{\tau_1, \tau_2, \tau_3, \tau_4\}$), one may define
four such hypersurfaces, the intersection of which is (generically) a
single point in an appropriate region of a well-behaved spacetime
geometry. Locally, points in such regions are distinguished by different
values of proper time broadcasts; the collection of proper times
$\underline{\tau}$ broadcasted by the four satellites may then be used
as coordinates in certain regions of spacetime.

A central problem in relativistic positioning system is that of
transforming between emission coordinates $\underline{\tau}$ and a more
standard coordinate system in a given spacetime geometry. If the
ephemerides of the satellites are known in a standard coordinate system
(Cartesian coordinates for flat spacetime, for instance), then the
emission coordinates $\underline{\tau}$ may be converted into the
coordinates for the emission points $X_I$ (which we collectively write
as $\underline{X}=\{X_1, X_2, X_3, X_4\}$), or the spacetime positions
of the satellites at the moments when the broadcasted values
$\underline{\tau}$ were emitted. To perform the coordinate
transformation, one must find in the standard coordinate system the
coordinates for the intersection point $X_{\rm c}$ of the future light
cones of four emission points $\underline{X}$, assuming a unique point
$X_{\rm c}$ exists in some appropriate region of spacetime. We refer to
the problem of finding the coordinates $X_{\rm c}$, given the
coordinates of the emission points $\underline{X}$ as the relativistic
location problem.

In Cartesian coordinates $t,x,y,z$ on flat spacetime, the coordinates
for the intersection point $X_{\rm c}$ must satisfy the following
constraint, which can in principle be solved using root-finding methods
in a brute-force approach:
\begin{equation} \label{IntersectionConstraint}
  \left( X_I^\mu - X_{\rm c}^\mu \right) 
  \left( X_I^\nu - X_{\rm c}^\nu \right)
  \eta_{\mu \nu} = 0,
\end{equation}
where here, $I \in \{1,2,3,4\}$, and $\eta_{\mu \nu}$ are the components
of the Minkowski metric:
\begin{equation} \label{MinkowskiMetric}
  \eta =
  \left[
  \begin{array}{cccc}
    -1  &  0  &  0  &  0 \\
     0  &  1  &  0  &  0 \\
     0  &  0  &  1  &  0 \\
     0  &  0  &  0  &  1 \\
  \end{array}
  \right] .
\end{equation}
In flat spacetime, several methods for computing such points, which
avoid brute-force root-finding methods, may be found in the literature,
for instance \cite{Coll:2009tm,Coll:2012dk} (implemented in
\cite{Puchades:2011bz}) and \cite{Cadez2010,Kostic:2015cca}.

%
%
\subsection{Transformation algorithm}\label{sec:IIB} 
\begin{figure}[]
  \includegraphics[width=1.0\columnwidth]{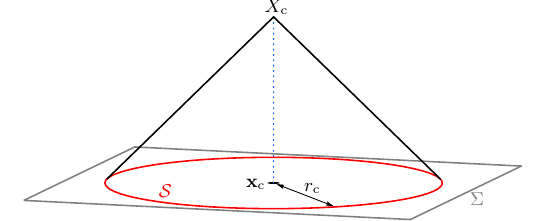}
  \caption{ $2+1$ illustration for the relativistic location algorithm
    in the case of a spacelike configuration hyperplane $\Sigma$ in an
    adapted frame. If the configuration hyperplane $\Sigma$ is
    spacelike, then one can perform a Lorentz transformation so that the
    emission points lie on a surface of constant time coordinate
    $t=X^{\prime 0}$ (the corresponding $t$ axis is vertical) with a
    value $t=t_0$. In this frame, the emission points lie on a sphere
    $\mathcal S$ (represented here as a circle) in the configuration
    hyperplane $\Sigma$. The problem consists of first finding the
    circumcenter ${\bf x}_{\rm c}$ and the circumradius $r_{\rm c}$. In
    this frame, the time it takes for light to travel the distance
    $r_{\rm c}$ is $\Delta t = r_{\rm c}$ in units where the speed of
    light is unity. The intersection point is then given by $X_{\rm c} =
    \left(t_0 + \Delta t,{\bf x}_{\rm c}\right)$. }
  \label{fig:Cones_Single}
\end{figure}

\begin{figure}[]
  \includegraphics[width=1.0\columnwidth]{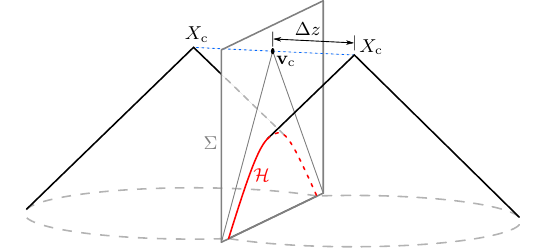}
  \caption{ $2+1$ illustration for the relativistic location algorithm
    in the case of a timelike configuration hyperplane $\Sigma$ in an
    adapted frame (as in Fig. \ref{fig:Cones_Single}, the $t$ axis is
    vertical). If $\Sigma$ is timelike, then one can perform a Lorentz
    transformation so that the emission points lie on a surface of
    constant coordinate $z=X^{\prime 3}$, with the value $z=z_0$. In
    this frame, the emission points lie on a hyperboloid $\mathcal H$
    (represented here as a hyperbola). The problem consists of first
    finding the coordinates of the vertex ${\bf v}_{\rm c}$ in this
    frame and the distance $R$, the latter being the distance between
    the vertex ${\bf v}_{\rm c}$ and a point on $\mathcal H$. The
    distance between the $z=z_0$ plane and the intersection point
    $X_{\rm c}$ is given by $R=\Delta z$. With these quantities in hand,
    one may obtain the intersection point $X_{\rm c} = \left({\bf
    v}_{\rm c},z_0\pm\Delta z\right)$. }
  \label{fig:Cones_Dual}
\end{figure}

Here, we describe
an algorithm for computing the intersection point $X_{\rm c}$ from four
emission points based on Lorentz transformations. To our knowledge, this
algorithm has not been explicitly described in the literature before,
though some of the methods may in principle be inferred from the
diagrams presented in \cite{Coll:2012dk} (which we reproduce here in
Figs. \ref{fig:Cones_Single} and \ref{fig:Cones_Dual}). Since this
algorithm distinguishes geometrically different configurations of the
emission points, it is physically intuitive and of conceptual utility,
and worth describing in detail here. Additionally, although this
algorithm is not the most optimal one for the flat spacetime case, our
implementation of it yields an improvement over a straightforward
implementation of the formula of \cite{Coll:2009tm,Coll:2012dk}
discussed below.

The algorithm we describe requires that the emission points
$\underline{X}$ are spacelike separated, or that
\begin{equation} \label{SpacelikeSep}
  \left( X_I^\mu - X_J^\mu \right) \left( X_I^\nu - X_J^\nu \right)
  \eta_{\mu \nu} \geq 0 ,
\end{equation}
for all $I$, $J$. The frame in which the emission points $\underline{X}$
are defined will be called $\mathcal{A}$. From these points, one may
construct three spacelike vectors $E_1$, $E_2$, $E_3$ in the following
manner:
\begin{equation} \label{Frame}
\begin{aligned}
E_1 &= X_2 - X_1 , \\
E_2 &= X_3 - X_1 , \\
E_3 &= X_4 - X_1 .
\end{aligned}
\end{equation}
These three vectors span a hyperplane $\Sigma$, called the configuration
hyperplane; from these three vectors, one may construct a vector normal
to the configuration hyperplane $\Sigma$ in the following manner
($\epsilon_{\nu \alpha \beta \delta}$ being the Levi-Civita tensor):
\begin{equation} \label{NormalRaw}
\begin{aligned}
N^\mu = \eta^{\mu \nu} \, \epsilon_{\nu \alpha \beta \delta}
        E^\alpha_1 E^\beta_2 E^\delta_3 ,
\end{aligned}
\end{equation}
and a unit normal vector:
\begin{equation} \label{NormalUnit1}
\begin{aligned}
n^\mu = \frac{q}{\sqrt{N_\sigma N^\sigma}} N^\mu ,
\end{aligned}
\end{equation}

\noindent where $q=\pm 1$, with the sign specified by the requirement
that $n$ be future pointing if timelike. 

\subsubsection{Spacelike configuration hyperplane}
We first consider the case where the configuration hyperplane $\Sigma$
is spacelike, so that the normal vector is timelike. In this case, one
may write:
\begin{equation} \label{NormalUnit2}
n = \left( \gamma , \beta \hat{r} \right) ,
\end{equation}
where $\hat{r} = (\hat{r}_x,\hat{r}_y,\hat{r}_z)$ is a unit vector.
$\gamma := n^0$ and $\beta = \sqrt{1-1/\gamma^2}$. One may then perform
a Lorentz transformation to a frame in which the spatial components of
the unit normal $n$ vanish. The Lorentz transformation matrix takes the
form
\begin{equation} \label{LorentzTransformation}
\Lambda =
\left[
\begin{array}{cccc}
   \gamma &  -\beta \gamma \hat{r} \\
   -\beta \gamma \hat{r}  &  (I + (\gamma-1) \, \hat{r} \otimes \hat{r})
    \\
\end{array}
\right] ,
\end{equation}
where $I$ is the identity matrix and $\otimes$ denotes a tensor product.

Since the Lorentz transformation $\Lambda$ transforms to a frame in
which $n^\prime = \Lambda \cdot n$ has no spatial components, it follows
that since the vectors $E_1$, $E_2$, and $E_3$ are orthogonal to $n$,
the time component of their transformed counterparts $E^\prime_1$,
$E^\prime_2$, and $E^\prime_3$ must vanish. From Eq. \eqref{Frame}, it
follows that the time components of the transformed emission points
$\underline{X}^\prime=\{X^\prime_1, X^\prime_2, X^\prime_3,
X^\prime_4\}$ are all equal; in this frame, the emission points all lie
on the same constant time slice $t=t_0$. The primed frame will be called
$\mathcal{B}$.

The problem of finding the intersection point of the light cones from
four points is simply a matter of finding the point spatially
equidistant from the four emission points. To see this, consider four
signals emitted from four points at the same instant. We seek the
spatial point at which the signals simultaneously arrive. The preceding
analysis establishes that one can find a reference frame (frame
$\mathcal{B}$) where the four emission points all lie on the same time
slice. Since the speed of light is constant in all frames, the point
where the signals simultaneously arrive must be spatially equidistant
from the four emission points. If the distance between the simultaneous
arrival point and each of the emission points is $r_{\rm c}$, then the
time coordinate in frame $\mathcal{B}$ is given by the time it takes for
light to travel a distance $r_{\rm c}$ (which has a value $r_{\rm c}$ in
units where the speed of light is $c=1$).

In a three-dimensional Euclidean space, this is a straightforward task.
Generically, four points that do not all lie in the same plane form the
corners of a tetrahedron. It is well known that, for any tetrahedron,
one can construct a circumsphere $\mathcal S$ that passes through all
the corners of a tetrahedron. The coordinates of the circumcenter
$\textbf{x}_{\rm c}$ specify the spatial coordinates of the intersection
point in frame $\mathcal{B}$, and the circumradius determines the time
coordinate. Given four (spatial) points
$\{\textbf{x}_1,\textbf{x}_2,\textbf{x}_3,\textbf{x}_4\}$, one can
compute the coordinates of the circumcenter $\textbf{x}_{\rm c}$ using
the following formulas \cite{LevyLiu2010}:
\begin{equation} \label{circumcenter1}
\textbf{x}_{\rm c} = \textbf{A}^{-1} {\bf u} ,
\end{equation}
where the $3 \times 3$ matrix $\textbf{A}$ and the vector ${\bf u}$ are
defined as
\begin{equation} \label{circumcenter2}
\textbf{A} :=
\left[
\begin{array}{c}
\,[\textbf{x}_2-\textbf{x}_1]^T \\
\,[\textbf{x}_3-\textbf{x}_1]^T \\
\,[\textbf{x}_4-\textbf{x}_1]^T
\end{array}
\right]
\qquad
{\bf u} := \frac{1}{2}
\left[
\begin{array}{c}
  \textbf{x}_2^2-\textbf{x}_1^2 \\
  \textbf{x}_3^2-\textbf{x}_1^2 \\
  \textbf{x}_4^2-\textbf{x}_1^2
\end{array}
\right].
\end{equation}
The circumradius $r_{\rm c}$ may then be computed using the formula:
\begin{equation} \label{circumradius}
r_{\rm c} = |\textbf{x}_{\rm c}-\textbf{x}_I|^{1/2}.
\end{equation}
The intersection of light cones in the frame $\mathcal{B}$ is then given
by:
\begin{equation} \label{BframeXc}
  X^\prime_{\rm c} = \left(t_0 + r_{\rm c},\textbf{x}_{\rm c}\right).
\end{equation}
To obtain the intersection of the light cones $X_{\rm c}$ in the
original frame $\mathcal{A}$, simply invert the Lorentz transformation:
\begin{equation} \label{AframeXc}
  X_{\rm c} = \Lambda^{-1} X^\prime_{\rm c} .
\end{equation}

There are instances in which this algorithm fails. For instance, the
algorithm may fail when the matrix $\bf{A}$ becomes degenerate, which
can occur if the emission points are collinear or coplanar (in which
case $r_{\rm c}$ diverges) \cite{Coll:2012dk}; these cases are discussed
in detail in \cite{AbelChaffee1991,ChaffeeAbel1994}.

\subsubsection{Timelike configuration hyperplane}
We now turn to the case in which the configuration hyperplane $\Sigma$
is timelike, which corresponds to a spacelike unit normal vector
$n^\mu$. In this case, the adapted frame is constructed differently; one
first performs a Lorentz transformation such that the spacelike $n^\mu$
is tangent to a surface of constant $t=X^{\prime 0}$ (here, $t=X^{\prime
0}$, $x=X^{\prime 1}$, $y=X^{\prime 2}$, $z=X^{\prime 3}$). Then, one
performs a spatial rotation so that $n^\mu$ is aligned with the $z$
axis. The configuration hyperplane $\Sigma$ in the resulting frame is
characterized by a constant $z=z_0$ coordinate. The points will lie on
an elliptic hyperboloid $\mathcal H$ formed by the intersection of the
past light cone for the solution point and the configuration hyperplane
$\Sigma$ (see Fig. \ref{fig:Cones_Dual}). The main task is to find the
coordinates for the vertex ${\bf v}_{\rm c} = \left(t_{\rm c},x_{\rm
c},y_{\rm c}\right)$ of the cone which the hyperboloid asymptotes to, as
well as the distance $R$ satisfying the following set of equations:
\begin{equation} \label{HyperboloidVertexEqs}
  \Phi_I= - R^2 ,
\end{equation}
where
\begin{equation} \label{HyperboloidPhiDef}
\begin{aligned}
  \Phi_I := \left(x_I - x_{\rm c}\right)^2 
            + \left(y_I - y_{\rm c}\right)^2 
            - \left(t_I - t_{\rm c}\right)^2 .
\end{aligned}
\end{equation}
By eliminating $R$, one may write this as a set of three equations:
\begin{equation} \label{HyperboloidVertexSystem}
\begin{aligned}
& \Phi_1 = \Phi_2 \\
& \Phi_2 = \Phi_3 \\
& \Phi_3 = \Phi_4,
\end{aligned}
\end{equation}

\noindent which can then be solved for the vertex coordinates ${\bf
v}_{\rm c}$ by way of a computer algebra system (we use {\it
Mathematica} \cite{Mathematica} to obtain explicit expressions).

The vertex coordinates provide the $\left(t,x,y\right)$ coordinates for
the intersection of future pointing light cones. The $z$ coordinate for
the intersection of light cones is given by
\begin{equation} \label{z-coord}
  z_{\rm c} = z_0 \pm R .
\end{equation}
In this case, one does not have a unique point for the intersection of
light cones---this is the bifurcation problem, which is discussed in
detail in \cite{Coll:2012dk}.

%
%
\subsection{Unified four emission point formula}\label{sec:IIC} 

It is also possible to calculate the intersection point $X_{\rm c}$
using a closed-form formula that applies regardless of the geometrical
configuration of the four emission points. Such a formula was presented
in \cite{Coll:2009tm,Coll:2012dk} and it is given by
\begin{equation} \label{CFM}
\begin{aligned}
X_{\rm c}^\mu = & X_1^\mu + y_*^\mu 
- \frac{y^\nu_*y_{*\nu} N^\mu}
  {y^\sigma_* N_\sigma\pm
  \sqrt{
    |(y^\sigma_* N_\sigma)^2 - y^\sigma_*y_{*\sigma} N^\tau N_\tau)|
    }
  } ,
\end{aligned}
\end{equation}

\noindent where $X_1^\mu$ is one of the emission points, $N^\mu$ is
defined in Eq. \eqref{NormalRaw}, and $y_*^\mu$ is given by
\begin{equation} \label{CFMy}
y_*^\mu := \frac{1}{\xi^\nu N_\nu} (i_\xi H)^\mu.
\end{equation}

\noindent Here, $\xi$ is any vector satisfying $\xi^\mu N_\mu \neq 0$,
and $i_\xi H$ is the interior product of $\xi$ and the two form $H$
[explicitly $(i_\xi H)^\mu = \eta^{\mu \nu} \xi^\sigma H_{\sigma \nu}$].
Given the definition for the frame vectors in Eq. \eqref{Frame}, the
two-form $H$ may be expressed as:
\begin{equation} \label{CFMH}
\begin{aligned}
&
H := \Omega_1 h(E_2,E_3) + \Omega_2 h(E_3,E_1) 
      + \Omega_3 h(E_1,E_2) , \\
&
\Omega_i := \eta_{\mu\nu} E^\mu_i E^\nu_i, 
\qquad
h(U,V)_{\alpha \beta} 
:=-\varepsilon_{\alpha \beta \mu \nu} U^\mu V^\nu ,
\end{aligned}
\end{equation}

\noindent with no sum on $i$ in the expression for $\Omega_i$.

Note that the $\pm$ sign in \eqref{CFM} means that there are always two
candidate solutions, even if the geometrical configuration leads to just
one. This complicates a straightforward use of the formula. The
determination of the correct sign is somewhat involved---especially when
the bifurcation problem mentioned in the previous section is present
\cite{Coll:2012dk}.

%
%
\subsection{Five emission points} \label{sec:IID} The bifurcation
problem may be solved most easily by including an additional point. If
five emission points are available, then one can obtain the intersection
point in a straightforward way. Following \cite{Ruggiero:2022}, one
begins with the constraint function:
\begin{equation} \label{IntersectionConstraintFunction}
  \Psi_I := \left( X_I^\mu - X_{\rm c}^\mu \right) 
            \left( X_I^\nu - X_{\rm c}^\nu \right)
  \eta_{\mu \nu},
\end{equation}

\noindent where now $I \in \{1,2,3,4,5\}$. One can take the differences
to form four unique equations of the following form:
\begin{equation} \label{IntersectionConstraintFunctionDiffs}
  \Psi_I - \Psi_J = \eta_{\mu \nu} 
              \left[ 2 \, X_{\rm c}^\mu \left(X_J^\nu-X_I^\nu\right) 
                    + X_I^\mu \, X_I^\nu - X_J^\mu \, X_J^\nu \right],
\end{equation}

\noindent which are linear in $X_{\rm c}^\mu$; one can reduce this to a
straightforward linear algebra problem, provided that the matrix
$X_J^\nu-X_I^\nu$ of emission point differences is nondegenerate. One
may observe that the matrix of emission point differences becomes
degenerate when any two emission points are brought together---for this
reason, one might encounter a loss of precision for closely separated
emission points. Though this formula requires an additional emission
point, it is preferred due to its computational simplicity and accuracy.

%
%
\subsection{Implementation, evaluation, and discussion} \label{sec:IIE}

The new algorithms we have presented here, as well as the formulas
described in \cite{Coll:2009tm,Coll:2012dk} and the five-point algorithm
of \cite{Ruggiero:2022} [which we have described in Eqs.
\eqref{IntersectionConstraintFunction} and
\eqref{IntersectionConstraintFunctionDiffs}], have been implemented in
the \textsc{cereal.jl} code,\footnote{The name is derived from the
pronunciation of the acronym SRL for special-relativistic locator.}
available at \cite{FHC2022srl}. We have written \textsc{cereal.jl} to
accommodate abstract datatypes; this allows user-specified floating
point precision. In the tests we perform, we consider two types of
floating point variables, the default \textsc{Float64} double precision,
and the \textsc{Double64} ``double double'' precision variables
implemented in the \textsc{DoubleFloats} library
\cite{DoubleFloats2018}.

Included in \textsc{cereal.jl} are test routines that perform tests of
the code by stochastically generating a set of $N$ emission points on
the past light cone of some intersection point $X_{\rm c}$, and
comparing the results $X_{\rm r}$ generated by the algorithms in
\textsc{cereal.jl} with the true value for $X_{\rm c}$. The points
$X_{\rm c}$ and $X_{\rm r}$ are compared according to Euclidean $L_2$
norms (with $|V|=\sqrt{V \cdot V}$ for some vector $V$):
\begin{equation} \label{Test}
  \varepsilon =
  |X_{\rm r} - X_{\rm c}|/|X_{\rm c}|
\end{equation}

\noindent In our tests, the most accurate algorithm is the five-point
formula of \cite{Ruggiero:2022}, which for $10^6$ test cases satisfies
$\varepsilon < 10^{-9}$ with double precision (\textsc{Float64}), and
$\varepsilon < 10^{-23}$ with extended precision (\textsc{Double64}).
The accuracy of the new algorithm presented in Sec. \ref{sec:IIB} and
the formula in Sec. \ref{sec:IIC} \cite{Coll:2009tm,Coll:2012dk} are
comparable to each other, but both are less accurate than the formula of
\cite{Ruggiero:2022}. For $10^6$ test cases, the errors for the
four-point methods in Secs. \ref{sec:IIB} and \ref{sec:IIC} typically
satisfy $\varepsilon < 10^{-5}$ with double precision
(\textsc{Float64}), and $\varepsilon < 10^{-14}$ with extended precision
(\textsc{Double64}).

Execution times differ greatly between the algorithms. On a standard
desktop computer (with an Intel i5-7500 processor), the five-point
formula of \cite{Ruggiero:2022} typically performs the computation in
$<1.5 ~\rm \mu s$. The four-point algorithms that we have implemented in
\textsc{cereal.jl} are significantly slower, despite only requiring four
emission points. For four emission points, our implementation of the
formula in \cite{Coll:2009tm,Coll:2012dk} typically requires $\sim 150
~\rm \mu s$ to perform the computation. The algorithm we have presented
here has improved performance, requiring a computation time of $\sim 36
~\rm \mu s$. Since the five-point formula is faster and yields results
with significantly higher accuracy, we employ it when computing the
initial guess for the curved spacetime algorithm that we will describe
in the next section.

We note that the algorithms described here may be used in conformally
flat spacetimes, since flat spacetimes and conformally flat spacetimes
share the same null cone and null geodesic structure on regions where
the conformal factor remains nonsingular. In particular, the
intersection point for four null cones in a conformally flat spacetime
will be the same as that for the underlying flat spacetime (underlying
in the sense that the metric for the conformally flat spacetime differs
from the flat spacetime by a conformal factor). This class of spacetimes
include cosmological spacetimes, such as de Sitter, anti-de Sitter and
the more general Friedmann-Lemaitre-Robertson-Walker spacetimes.


%
%

%
%
\section{Relativistic location in curved spacetime}\label{sec:III}

%
%
\subsection{Geodesics}
For general spacetime geometries, described by a metric tensor $g_{\mu
\nu}$ and its inverse $g^{\mu \nu}$, the problem of finding the
intersection point $X_{\rm c}$ of four future pointing light cones
(provided that such a point exists) amounts to finding the intersection
of four null geodesics from the emission points $\underline{X}$; this
follows from the fact that for some emission point $X_I$, a point
$X_{\rm p}$ in the future pointing null cone lies on a geodesic
connecting $X_{\rm p}$ and $X_I$. Note also that the emission points
$\underline{X}$ lie on the past light cone of $X_{\rm c}$. Given some
inverse metric $g^{\mu \nu} = g^{\mu \nu}(x)$ describing the spacetime
geometry, an affinely parametrized null geodesic may be described by the
Hamiltonian
\begin{equation} \label{Hamiltonian}
  H := \frac{1}{2} \, g^{\mu \nu} \, p_\mu \, p_\nu ,
\end{equation}

\noindent where the four-momenta are given by
\begin{equation} \label{FourMomenta}
  p_\mu = g_{\mu \nu} \frac{dx^\nu}{d\lambda} ,
\end{equation}

\noindent and the associated Hamilton equations are
\begin{equation} \label{HamiltonianEquations}
  \begin{aligned}
  \frac{d{x}^\mu}{d\lambda} = \frac{\partial H}{\partial p_\mu} ,
  \qquad \qquad
  \frac{d{p}_\mu}{d\lambda} = - \frac{\partial H}{\partial x^\mu} .
  \end{aligned}
\end{equation}

\noindent For null geodesics, the initial data at $\lambda=0$ is given
by an initial point ${x}_0^\mu$ and an initial three velocity ${\bf
v}^i$, with the initial four-momentum $p_\mu |_{\lambda=0}$ satisfying
the following (with $i \in \{1,2,3\}$):
\begin{equation} \label{HamiltonianEquationsICs}
  \begin{aligned}
  \left.\delta{^i}{_\mu} \frac{dx^\mu}{d\lambda}\right|_{\lambda=0} 
  = {\bf v}^i,
  \qquad
  \left. g_{\mu \nu}({x}_0) 
  \frac{dx^\mu}{d\lambda}\frac{dx^\nu}{d\lambda} \right|_{\lambda=0} 
  = 0.
  \end{aligned}
\end{equation}

\noindent The solution to Hamilton's equations is formally given by
$x^\mu=x^\mu(\lambda,{x}_0,{\bf v})$. Since $\lambda$ is an affine
parameter, one can redefine $\lambda$ up to linear transformations---it
is therefore always possible to rescale $\lambda$ so that it takes
values in the domain $\lambda \in [0,1]$, with $\lambda=1$ being the
final point.

%
%
\subsection{Geodesic intersection}
The problem of finding the intersection of light cones in a slightly
curved spacetime may be reformulated in terms of null geodesics
\cite{Carloni:2018cuf}. Consider four formal solutions to Hamilton's
equations \eqref{HamiltonianEquations}, distinguished by the indices
$I\in\{1,2,3,4\}$, that have end points ${\rm x}^\mu_I$ which are
functions of the initial data ${X}_I$ and ${\bf v}_I$:
\begin{equation} \label{GeodesicSolns}
  {\rm x}^\mu_I = {\rm x}^\mu_I({X}_I,{\bf v}_I) 
  = x^\mu_I(1,{X}_I,{\bf v}_I) .
\end{equation}

\noindent Then define the following vector valued function:
\begin{equation} \label{ZeroFunction}
  \begin{aligned}
  F := \left( {\rm x}_1 - {\rm x}_2 , {\rm x}_1 - {\rm x}_3 
            , {\rm x}_1 - {\rm x}_4 \right).
  \end{aligned}
\end{equation}

\noindent where $F=F(\underline{X},v)$ [with $v=({\bf v}_1,{\bf
v}_2,{\bf v}_3,{\bf v}_4)$]. Observe that upon evaluation, the function
$F$ yields a 12 component vector. The intersection of four null
geodesics is given by the condition 
\begin{equation} \label{ZeroFunction0}
  F(\underline{X},v) = 0 .
\end{equation}

\noindent The problem of solving the system of 12 equations in Eq.
\eqref{ZeroFunction0} is a standard root-finding problem. In particular,
given a set of four emission points $\underline{X}=\{X_1,X_2,X_3,X_4\}$,
one solves Eq. \eqref{ZeroFunction0} for the 12 quantities $v=({\bf
v}_1,{\bf v}_2,{\bf v}_3,{\bf v}_4)$ that constitute the initial data.

%
%
\subsection{Initial data}
From here on, we write $f(v)=F(\underline{X},v)$ for simplicity,
suppressing the dependence on emission points $\underline{X}$. The
specific root finding algorithm we intend to employ will be based on an
iterative quasi-Newton method, which requires an initial guess. It is
therefore appropriate to begin by assuming that the spacetime geometry
is slightly curved; the flat spacetime algorithms described earlier may
then be used to construct an initial guess for $v$.

Initial data for the geodesics is constructed from the emission points
$\underline{X} = \{X_1, X_2, X_3, X_4\}$ and the flat spacetime
intersection point $X_{\rm c}$. From these, one obtains the initial
guess for the vector $v=({\bf v}_1,{\bf v}_2,{\bf v}_3,{\bf v}_4)$:
\begin{equation} \label{boldvinitialguess}
  {\bf v}^i_I := X^i_{\rm c} - X^i_I,
\end{equation}

\noindent in units of dimensionless affine parameter. From $v$ and
$\underline{X}$, one may construct the initial data for the geodesics 
by first constructing the vector $V_I$ for each geodesic:
\begin{equation} \label{initialV}
  V_I = \left( V^0_I , {\bf v}^1_I , {\bf v}^2_I , {\bf v}^3_I \right) ,
\end{equation}

\noindent where $V^0_I$ is determined by the condition
\begin{equation} \label{nullV}
  V^\mu_I \, V^\nu_I \, g_{\mu \nu}(X_I) = 0 .
\end{equation}

\noindent The conjugate momenta are given by
\begin{equation} \label{momenta}
  p^{I}_\mu = V^\nu_I \, g_{\mu \nu}(X_I) .
\end{equation}

\noindent The initial positions $X_I$ and initial conjugate momenta
$p^{I}$ provide initial data for Eq. \eqref{HamiltonianEquations}, which
may then be solved to compute the value of $f(v)$ according to Eq.
\eqref{ZeroFunction}.

%
%
\subsection{Root finding}
In general, one does not possess analytical solutions to the geodesic
equation \eqref{HamiltonianEquations} for a generic metric $g_{\mu
\nu}$. To evaluate the function \eqref{ZeroFunction}, one must therefore
solve the geodesic equation \eqref{HamiltonianEquations} numerically for
each emission point. One might expect a root finding algorithm for Eq.
\eqref{ZeroFunction0} to be computationally expensive, particularly in
the computation of a Jacobian.

However, libraries for efficiently computing the Jacobian of generic
functions have become available in recent years, in particular those
that employ automatic differentiation methods. Automatic differentiation
refers to a set of methods which, by way of the chain rule, exploit the
fact that all numerical computations can in principle be broken down
into finite compositions of elementary arithmetic operations. These
methods can in principle be used to numerically compute the derivatives
of programs to machine precision with a minimal computational overhead.
A detailed discussion of automatic differentiation may be found in
\cite{neidinger2010,baydin2018}. In our approach, we obtain the Jacobian
of $=F(\underline{X},v)$ by automatic differentiation of numerical
solutions to the geodesic equation in a generic slightly curved
spacetime.\footnote{We note that automatic differentiation methods have
been previously proposed for reducing the computational complexity for
relativistic location in the Schwarzschild spacetime
\cite{Delva:2009tk}, and we also note that, in
\cite{burton2007numerical}, automatic differentiation methods have been
proposed as a way to obtain Taylor expansions of the initial value
problem for the geodesic equation.}

The specific root finding algorithm we employ is based on an iterative
quasi-Newton Broyden method \cite{Broyden1965,Press2007NR}, which we
summarize here. The task at hand is to obtain the root of some function
$f(v)$. In the initial iteration, the Jacobian of $f(v)$ is computed
using automatic differentiation methods. We also employ automatic
differentiation in computing the gradient of the Hamiltonian, a strategy
also employed in \cite{Christian:2020xrp} for solving the geodesic
equation in Hamiltonian form. Given the Jacobian ${\bf J}$ and its
inverse ${\bf J}^{-1}$, at some iteration ${\rm i}$, one can update $v$
according to the Newton prescription:
\begin{equation} \label{NewtonUpdated}
  \begin{aligned}
  v_{{\rm i}+1} = v_{\rm i}
            +
            {\bf J}^{-1}_{{\rm i}} f(v_{\rm i}) .
  \end{aligned}
\end{equation}
In the standard Broyden method (alternatively referred to as the
``good'' Broyden method), the first iteration is given by Eq.
\eqref{NewtonUpdated}, with the Jacobian computed by differentiation.
For the subsequent iterations, one computes the following:
\begin{equation} \label{Deltavf}
  \begin{aligned}
  \Delta v_{\rm i} &= v_{\rm i} - v_{{\rm i}-1} \\
  \Delta f_{\rm i} &= f(v_{\rm i}) - f(v_{{\rm i}-1}) ,
  \end{aligned}
\end{equation}
The inverse Jacobian ${\bf J}^{-1}$ is then updated according to the
Sherman-Morrison formula:
\begin{equation} \label{ShermanMorrison}
  \begin{aligned}
    {\bf J}^{-1}_{{\rm i}+1} = {\bf J}^{-1}_{\rm i}
                +
                \frac{\Delta v^{T}_{\rm i} 
                      - {\bf J}^{-1}_{\rm i} \, \Delta f_{\rm i} 
                     }
                     {\Delta v^{T}_{\rm i} \, {\bf J}^{-1}_{\rm i} \, 
                      \Delta f_{\rm i} 
                     }
                \Delta v^{T}_{\rm i} \, {\bf J}^{-1}_{\rm i}.
  \end{aligned}
\end{equation}

\noindent One may then use Eq. \eqref{ShermanMorrison} in conjunction
with \eqref{NewtonUpdated} to iteratively solve for the root of $f(v)$.
The termination of the algorithm is determined by the behavior of
$f_{\rm i}$; if a local minimum is detected within a specified range of
iterations, the algorithm terminates and the results corresponding to
the minimum are returned. In case the algorithm does not converge, a
hard termination limit is used.

Given a root for $f(v)$, one can obtain the intersection point by
solving the geodesic equations once more with the updated values for the
initial data constructed from $v$ and $\underline{X}$, and averaging
over the end points (which are assumed to be close).

%
%
\subsection{The squirrel algorithm}
We now summarize the curved spacetime algorithm employed in the
\textsc{squirrel.jl} code:\footnote{The name is derived from the
pronunciation of the acronym SCuRL for slightly curved relativistic
locator.}

\begin{enumerate}
  \item First, apply a flat spacetime algorithm (either that of Secs.
        \ref{sec:IIB} or \ref{sec:IID}) to the emission points
        $\underline{X}$ to obtain a guess for the intersection point and
        initial velocities.
  \item Apply a root finding algorithm to the function     
        $f(v)=F(\underline{X},v)$ to obtain the initial velocities $v$
        for subsets of four emission points.
  \item Integrate the geodesics with the resulting initial velocities
        $v$ and emission points $\underline{X}$ to find the intersection
        point.
\end{enumerate}
As indicated, steps 2 and 3 of the above algorithm are applied to a
subset of four emission points. If additional emission points are
available, an outlier algorithm, described in the next subsection, is
employed to exclude large errors. 

%
%
\subsection{Outlier detection}
There are instances in which the algorithm described in this section can
generate large errors, which can result from a combination of large
errors in the initial guesses provided by the flat spacetime algorithm
and convergence failures in the Broyden algorithm. One might expect such
errors to occur, since the function $F(\underline{X},v)$ is generally
nonlinear. To increase the reliability of the algorithm, we describe
here methods that can mitigate the effects of these errors when
additional emission points are available.

As discussed before, given $N>4$ emission points, one can choose up to
$C(N,4)$ combinations of four emission points $\underline{X}$, and for
each set $\underline{X}$, the previously described algorithm can be
applied to obtain a total of $C(N,4)$ intersection points. Since there
is only one receiver for the emission data, all $C(N,4)$ intersection
points should agree. If errors in the algorithm are assumed to be rare,
one can employ an outlier detection algorithm that can identify the
intersection points that strongly deviate from the others.

We employ a simple outlier detection algorithm, which begins by first
computing the median values for the intersection points, and then
computes the deviation of each intersection point from the median. The
points which deviate from the median beyond a user-specified threshold
are then discarded. The final intersection point is then computed from
the remaining intersection points.

%
%
\subsection{Remarks on implementation}
The algorithm described here is implemented in the \textsc{squirrel.jl}
code (available at \cite{FHC2022scurl}). The \textsc{squirrel.jl} code
is written in the Julia language, which is ideal for implementing the
squirrel algorithm due to the state of the art automatic differentiation
and ODE solver libraries available. Automatic differentiation is handled
using the \textsc{ForwardDiff.jl} forward-mode automatic differentiation
library \cite{RevelsLubinPapamarkou2016}, and geodesics are integrated
using the recommended Verner seventh order Runge-Kutta integrator
\textsc{AutoVern7} \cite{verner2010numerically} in
\textsc{OrdinaryDiffEq.jl} \cite{DifferentialEquations.jl-2017}, which
features stiffness detection and automated switching to a specified
stiff integrator (we use the fifth order Rosenbrock method integrator
\textsc{Rodas5} \cite{DiMarzo1993}). Though our system is Hamiltonian,
we have avoided symplectic integrators in favor of integrators with
adaptive time stepping in order to minimize execution time.

The Broyden algorithm is implemented directly, depending only on
standard Julia libraries. The default termination limit is set to $24$.
The initial guess is provided by one of the flat spacetime algorithms
implemented in the \textsc{cereal.jl} code, depending on the number of
emission points available; if $N=4$ emission points are available, then
the flat spacetime algorithm presented in Sec. \ref{sec:IIB} is employed
(in which case, our implementation returns two points), but if $N \geq
5$ emission points are available, then the formula of
\cite{Ruggiero:2022} reviewed in Sec. \ref{sec:IID} is employed. The
outlier detection algorithm becomes active for $N \geq 5$ emission
points, and is applied to the location algorithm of \textsc{squirrel.jl}
to remove results with large errors.

Since there is now widespread availability of devices with
multithreading capabilities, the \textsc{squirrel.jl} code employs
multithreading on loops containing the integration of geodesics and the
automatic differentiation of geodesic solutions. With multithreading
enabled on a desktop computer with four cores and four threads (Intel
i5-7500), the \textsc{squirrel.jl} code can establish a position from
five emission points in under $1~{\rm s}$ for reasonably simple
spacetime geometries---benchmarks will be discussed in Sec.
\ref{sec:Benchmarks}.


%
%

%
%
\section{Curved spacetime geometries}\label{sec:IV}

%
%

\begin{figure}[]
  \includegraphics[width=1.0\columnwidth]
  {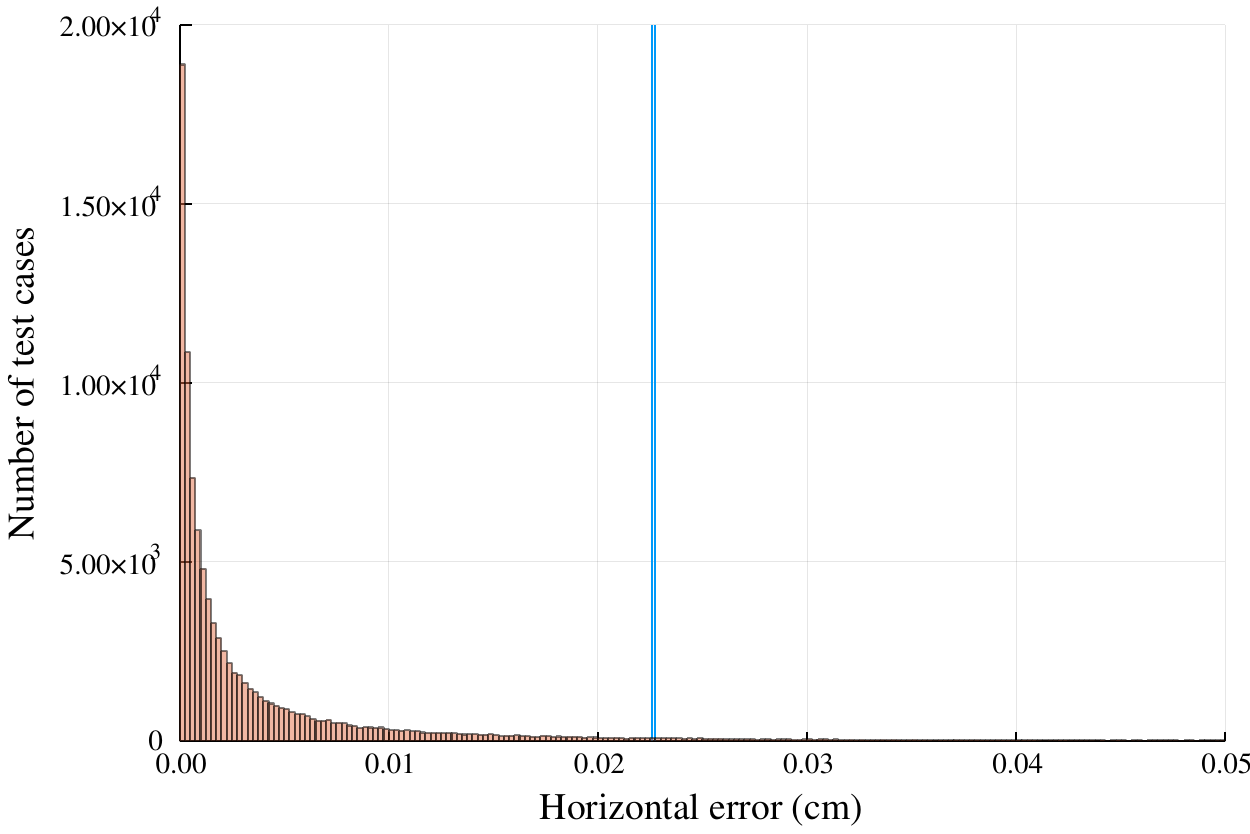}
  \caption{ Horizontal positioning errors for our implementation of the
  five emission point algorithm of \cite{Ruggiero:2022} relative to the
  Kerr-Schild geometry for $10^5$ test cases. The closely spaced
  vertical lines correspond to the rms values and $95\%$ confidence
  level for the errors, with respective values of $0.0226$ and
  $0.0227~{\rm cm}$. Out of $10^5$ samples, two samples ($0.002\%$) have
  an error $>2~{\rm cm}$ with the largest error being $3.3~{\rm cm}$. }
  \label{fig:plotxHerrKerr}
\end{figure}

\begin{figure}[]
  \includegraphics[width=1.0\columnwidth]
  {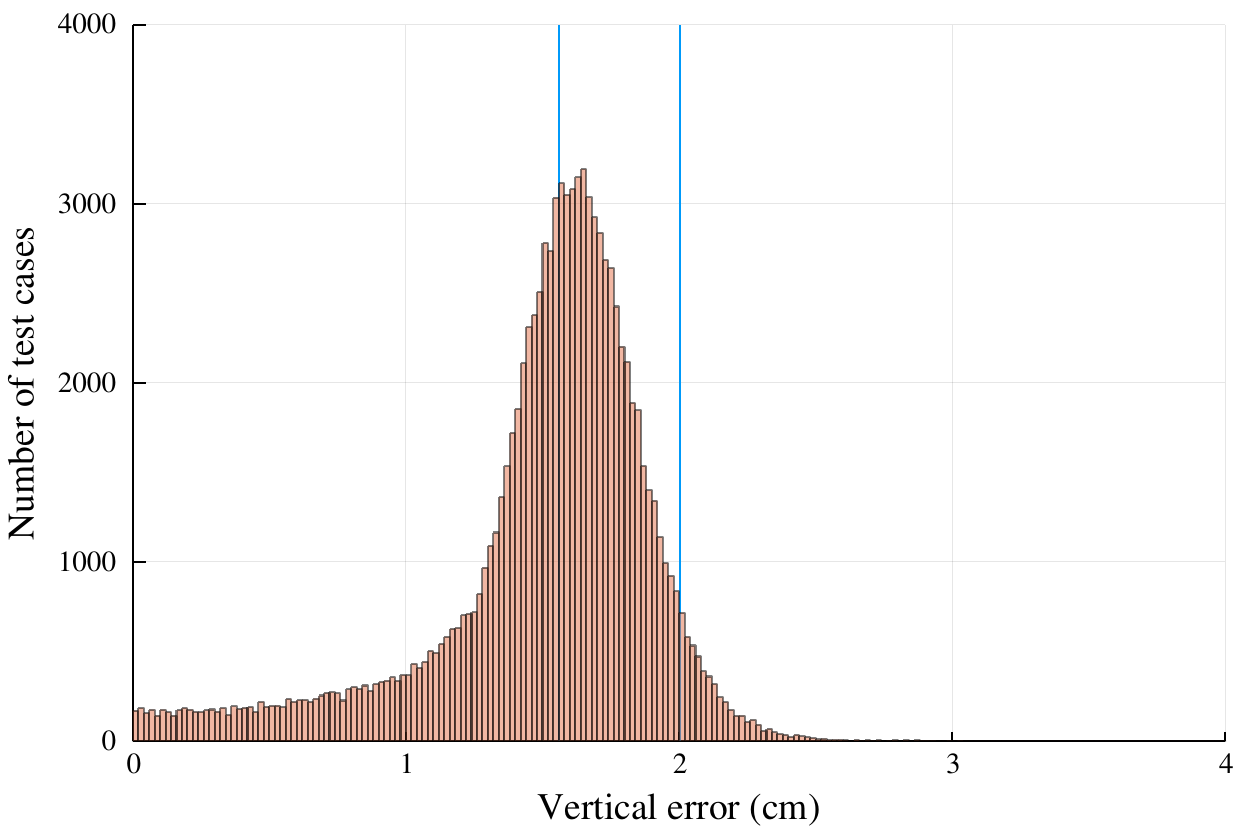}
  \caption{ Vertical positioning errors for our implementation of the
  five emission point algorithm of \cite{Ruggiero:2022} relative to the
  Kerr-Schild geometry for $10^5$ test cases. The vertical lines
  correspond to the rms values and $95\%$ confidence level for the
  errors, with respective values of $1.56$ and $2.00~ {\rm cm}$. Out of
  $10^5$ samples, $11$ samples ($0.011\%$) have an error $>5~{\rm cm}$
  with the largest error being $15.0~{\rm cm}$. }
  \label{fig:plotxVerrKerr}
\end{figure}

\subsection{The Kerr-Schild metric}\label{sec:MetricsKS} In this
section, we describe some specific choices for the spacetime metric
$g_{\mu \nu}$ used in our tests. In general relativity, the spacetime
geometry surrounding a stationary rotating object in a vacuum is given
by the Kerr geometry \cite{Kerr:1963ud}, described by the following
metric in Kerr-Schild coordinates \cite{Kerr:2009}:
\begin{equation} \label{KerrSchild}
  \begin{aligned}
  g_{\mu \nu} &= \eta_{\mu \nu} + f \, k_\mu \, k_\nu , \\
  k_\mu &= \left( 1 , \frac{r \, x + a \, y}{r^2+a^2}
                , \frac{r \, y - a \, x}{r^2+a^2} , \frac{z}{r}
                \right) , \\
  f &= \frac{2 \, G \, M \, r^3}{r^4+a^2 \, z^2} ,
  \end{aligned}
\end{equation}
where $G$ is the gravitational constant, $M$ is the mass, and $a$ is the
spin parameter. The radius $r$ may be compactly expressed as the
solution to
\begin{equation} \label{KerrSchildrad}
  \frac{x^2+y^2}{r^2+a^2} + \frac{z^2}{r^2} = 1 .
\end{equation}

At the surface of the Earth, the spacetime curvature is small, so one
might ask whether the flat spacetime algorithms suffice for positioning,
neglecting tropospheric and ionospheric corrections. To determine
whether this is indeed the case, we test the five-point algorithm in
Eqs. \eqref{IntersectionConstraintFunction} and
\eqref{IntersectionConstraintFunctionDiffs} against the Kerr geometry.
Recalling that emission points lie on the past light cone of the
intersection point $X_{\rm c}$, we stochastically generate intersection
points $X_{\rm c}$, then construct initial data for past directed null
geodesics. To obtain the emission points $\underline{X}=\{X_1, X_2, X_3,
X_4, X_5\}$, we integrate the geodesic Hamilton equations for five sets
of initial data for the null geodesics. The resulting emission points
$\underline{X}$ are used with the formula of \cite{Ruggiero:2022} to
obtain $X_{\rm c}$, the intersection point in flat spacetime. The
positioning error is given by (with vertical bars denoting the $L_2$
Euclidean distance norm)
\begin{equation} \label{CerealKSPositioningError}
  \epsilon_{\rm KSc} = |\hat{P}({\bf x}_{\rm tar}-{\bf x}_{\rm c})|.
\end{equation}

\noindent Here, $\hat{P}(\cdot)$ is a projection operator (projecting to
horizontal or vertical directions relative to some surface), ${\bf
x}_{\rm c}$ form the spatial components of $X_{\rm c}$, and ${\bf
x}_{\rm tar}$ denotes the spatial components of $X_{\rm tar}$ which is
the true intersection point for the future light cones of
$\underline{X}$ with respect to the Kerr-Schild metric.

We consider a Kerr-Schild metric with parameter choices $G M=1$ and $a =
738$ (the latter corresponding to the angular momentum for the Earth).
We perform a test with $10^5$ randomly generated target points $X_{\rm
tar}$ on the WGS-84 reference ellipsoid \cite{WGS84-2000} (setting $M$
to be the Earth mass) and randomly generated initial datasets for null
geodesics. The result, illustrated in Figs. \ref{fig:plotxHerrKerr} and
\ref{fig:plotxVerrKerr}, indicates that the error satisfies
$\epsilon_{\rm KSc} \leq 2~ {\rm cm}$ for $95\%$ of the points in the
vertical direction (the direction orthogonal to the reference
ellipsoid), and $\epsilon_{\rm KSc} \leq 3~ {\rm mm}$ in the horizontal
direction. In a vacuum, the five emission point algorithm in flat
spacetime suffices for positioning to an accuracy on the order of a
centimeter. This result is consistent with those of
\cite{Puchades:2016myk}, where it is also argued that the dominant
errors from spacetime curvature come from the determination of satellite
orbits, rather than the bending of photon trajectories.

%
%
\subsection{The Gordon metric}
If one seeks centimeter-scale accuracy in a vacuum on terrestrial
scales, then flat spacetime algorithms suffice. However, for terrestrial
positioning, tropospheric and ionospheric effects significantly affect
the propagation of electromagnetic signals and introduce errors in the
computed position. From a general relativistic perspective, one might be
tempted to dismiss tropospheric and ionospheric effects as ancillary
(practical considerations aside), as the underlying spacetime geometry
does not depend to a significant degree on tropospheric and ionospheric
profiles. However, in a positioning system based on the exchange of
electromagnetic\footnote{Signals encoded in weakly interacting particles
such as neutrinos may offer a possible alternative for relativistic
positioning that would avoid the need to consider tropospheric and
ionospheric effects.} signals, such effects will deform the emission
coordinates, and, in this sense, tropospheric and ionospheric effects
should still be taken into consideration even if one insists on a
fundamentally relativistic approach.

Fortunately, as indicated in \cite{Tarantola:2009hk}, the framework of
general relativity can by way of analog spacetime geometries incorporate
the effects of dielectric media on electromagnetic signal propagation.
In dielectric media, light propagation may under certain conditions be
described with the geodesics of the Gordon metric
\cite{Gordon1923,Pham1956,Barcelo:2005fc}, which has the form
\begin{equation} \label{GordonMetric}
  \begin{aligned}
    \bar{g}_{\mu \nu} &= g_{\mu \nu}
    + \left(1-\frac{1}{n^2}\right) u_\mu u_\nu ,
  \end{aligned}
\end{equation}

\noindent where $u^\mu$ corresponds to the four-velocity of the medium,
and $n$ is an effective index of refraction. To simplify the analysis,
we will neglect the rotation of the Earth (in general, the corotation of
the medium can be included through the four-velocity $u^\mu$). The
tropospheric index of refraction has a sea level value of $n_{\rm atm} -
1 \sim 2.7 \times 10^{-4}$, and the effective index of refraction for
the ionosphere has a maximum value on the order of $n_{\rm ion} - 1 \sim
4.0 \times 10^{-5}$. It follows that the tropospheric and ionospheric
corrections to the metric are of the respective orders $10^{-4}$ and
$10^{-5}$. In contrast, the difference $f \, k_\mu \, k_\nu$ between the
components of the Kerr-Schild metric and the Minkowski metric is roughly
on the order of $10^{-9}$ at the surface of the Earth, so tropospheric
and ionospheric effects dominate.

%
%
\subsection{Weak field metric}
At this point we emphasize that when performing tests with the analog
Gordon metric, we incorporate gravitational effects with the weak-field
metric, rather than the Kerr metric. The weak field metric has the form
\begin{equation} \label{WeakFieldMetric}
  g_{\mu \nu} = \eta_{\mu \nu} - 2 \, V \, \delta_{\mu \nu},
\end{equation}
where $V$ is the gravitational potential of the Earth, which takes the
form
\begin{equation} \label{Potential}
  V = -\frac{G M}{r} \left[1
      - J_2 \left(\frac{a_{\rm ell}^2}{r^2}P_2(\cos \theta)\right)
      \right],
\end{equation}
where $r^2=x^2+y^2+z^2$, $P_2$ is a Legendre polynomial of degree $2$,
$J_2$ is a quadrupole moment of the Earth, which takes a value of 
\cite{Ashby:2003vja}
\begin{equation} \label{QuadrupoleMoment}
  J_2 = 1.0826300 \times 10^{-3}.
\end{equation}
The quantity $a_{\rm ell}$ is the equatorial radius of the Earth, and is
one of the parameters of the reference ellipsoid, which approximates the
Earth's geoid up to roughly $100~{\rm m}$. Following
\cite{Ashby:2003vja}, the reference ellipsoid we use is the WGS-84
standard \cite{WGS84-2000}, which corresponds to the following values
for the semimajor axis $a_{\rm ell}$ and the semiminor axis $b_{\rm
ell}$:
\begin{equation} \label{WGS84EllipsoidCoeffs}
  \begin{aligned}
    a_{\rm ell} &=  6378.137~{\rm km}, \\
    b_{\rm ell} &=  6356.752314245~{\rm km}.
  \end{aligned}
\end{equation}


%
%

%
%
\section{Index of refraction models}\label{sec:V} 

We now turn to the construction of models for the effective index of
refraction, which we will use in the analog Gordon metric
\eqref{GordonMetric} for our tests of the algorithm. The effective index
of refraction is then given by the following expression
\begin{equation} \label{EffectiveIndexofRefraction}
  n_{\rm eff} = 1 + \Delta n_{\rm atm} + \Delta n_{\rm ion}.
\end{equation}
In this section, we describe the construction of simplified profiles for
$\Delta n_{\rm atm}$ and $\Delta n_{\rm ion}$ which we will use in
evaluating the \textsc{squirrel.jl} code.

%
%
\subsection{Atmospheric model}
Given the pressure and temperature profiles for the atmosphere, the
profile for the atmospheric index of refraction (here excluding
contributions from the ionosphere) can be computed from the revised
Edl{\'e}n equation \cite{Birch1993,*Edlen1966} for the refractive index
of air:
\begin{equation} \label{EdlenEquation}
  \begin{aligned}
  \Delta n_{\rm air} = \,
  & \frac{\Delta n_{\rm s} \left[P/{\rm Pa}\right]}{96095.43}
  \\
  & \times
  \frac{1 + 10^{-8}
  \left(0.601-0.00972\left[T/^{\circ}{\rm C}\right]\right)
  \left[P/{\rm Pa}\right]}
  {1+0.0036610 \left[T/^{\circ}{\rm C}\right]}
  ,
  \end{aligned}
\end{equation}

\noindent where $\Delta n_{\rm air} = n_{\rm air} - 1$, and $\Delta
n_{\rm s}$ is given by
\begin{equation} \label{dns}
  \begin{aligned}
  \Delta n_{\rm s} \times 10^8 = \, &
  8342.54
  +
  \frac{2406147}{130-1/\lambda^2}
  +
  \frac{15998}{38.9-1/\lambda^2}.
  \end{aligned}
\end{equation}

\noindent Following \cite{Vasylyev2019}, one may obtain standard
atmospheric temperature and pressure profiles from one of several
atmospheric models, for instance the U.S. Standard Atmosphere model
\cite{Atmosphere1976US}, or the more detailed NRLMSISE-00 model
\cite{Picone2002}. Using the former, atmospheric index of refraction
profiles up to $80~{\rm km}$ are computed, and we fit the computed
values to a function of the form:
\begin{equation} \label{IndexRefractionForm}
  \Delta n_{\rm atm} = \frac{A_1}{B_1 + C_1 \left(h-H_1)\right)}
  + \frac{A_2}{B_2 + C_2 \left([h]-H_2)\right)},
\end{equation}
with $h$ denoting altitude (in ${\rm km}$) from an appropriate reference
ellipsoid. The fitted parameter values are
\begin{equation} \label{IndexRefractionCoeffs1}
  \begin{aligned}
    A_1 &= -222.666,    &\qquad   A_2 &= -253.499,  \\
    B_1 &= 99.0621 ,    &\qquad   B_2 &= 112.757,  \\
    C_1 &= 0.157823 ~{\rm km}^{-1},
        &\qquad C_2 &= 0.179669 ~{\rm km}^{-1} ,\\
    H_1 &= -7.1541  ~{\rm km} ,
        &\qquad   H_2 &= -7.15654 ~{\rm km}.
  \end{aligned}
\end{equation}
Though it would be aesthetically preferable to employ exponential
functions in our model, we refrain from using them to avoid potential
instabilities in the libraries we used for the integration of ODEs.
Since the contributions from $\Delta n_{\rm atm}$ are concentrated in
the troposphere, the contributions from $\Delta n_{\rm atm}$ will be
referred to as tropospheric.

%
%
\subsection{Ionospheric model}
The effective index of refraction for electromagnetic wave propagation
the ionosphere is given by the Appleton–Hartree equation
\cite{Bittencourt2004,Appleton1932,*Hartree1931}. We consider here an
approximation which assumes a collisionless plasma, and signal
frequencies $\omega:=2\pi f$ much greater than the gyrofrequency
$\omega_{\rm g}:=|q {\bf b}_{\rm E} / m_{\rm e}|$ with ${\bf b}_{\rm E}$
being the Earth's magnetic field, and $q$, $m_{\rm e}$ being the
respective charge and mass of the electron. For GNSS signals, $f \sim
10^9~{\rm Hz}$, and $\omega_{\rm g}/\omega \approx 3 \times 10^{-3}$, so
this approximation is reasonable. The corrections to the effective index
of refraction $n_{\rm ion}$ from the gyrofrequency are in fact
proportional to $\omega_{\rm g}/\omega$, and depend on the angle between
the direction of radio wave propagation and ${\bf b}_{\rm E}$. If
gyrofrequency corrections become important, nongeometrical corrections
to the geodesic equation may be needed, in which case the Gordon metric
alone does not suffice for characterizing the propagation of
electromagnetic signals. However, one may nonetheless suppress such
corrections with higher signal frequencies.

Under the assumptions in the preceding paragraph, the Appleton-Hartree
formula for the ionospheric phase index of refraction $n_{\rm ph}$ may
be approximated as
\begin{equation} \label{IndexofRefractionIonospherePhase}
  \Delta n_{\rm ph} \approx - \frac{\omega_{\rm p}^2}{f^2} =
  - \left(4.024 \times 10^{-17}\right) [N_{\rm e}/{\rm m}^{-3}],
\end{equation}

\noindent where $\Delta n_{\rm ph} = n_{\rm ph}-1 $, and $\omega_{\rm
p}^2=q^2 N_{\rm e}/2 \epsilon_0 \, m_{\rm e} $ is the squared plasma
frequency, with $\epsilon_0$ being the vacuum permittivity, and $N_{\rm
e}$ the electron density in ${\rm m}^{-3}$. Since the corrections from
the gyrofrequency $\omega_{\rm g}$ are linear, we assume that the index
$\Delta n_{\rm ph}$ can only be modeled up to a precision of $0.3 \%$
for signals in the GHz range. It should be mentioned that since $n_{\rm
ph}<1$, $n_{\rm ph}$ can only be the index of refraction associated with
the phase velocity. To obtain the index of refraction associated with
the group velocity, one employs the dispersion relation $\omega^2 = k^2
+ \omega_{\rm p}^2$ for cold, collisionless plasmas
\cite{Bittencourt2004} to obtain the following expression for the group
index of refraction (with $\Delta n_{\rm ion} = n_{\rm ion}-1 $)
\begin{equation} \label{IndexofRefractionIonosphere}
  \Delta n_{\rm ion} \approx - \Delta n_{\rm ph}.
\end{equation}

The electron density $N_{\rm e}$ can be determined by measurement and
modeling; the Global Positioning System employs the Klobuchar model
\cite{klobuchar1987ionospheric} and the Galileo GNSS makes use of the
NeQuick-G ionospheric model detailed in \cite{EU2016ion} (which is a
revised version of the NeQuick model in \cite{radicella2001evolution}).
In these models, the ionospheric profile is described in terms of the
dimensionless Epstein function:
\begin{equation} \label{EpsteinFunction}
  Ep(h,h_{\rm c},B) := \frac{4 \exp \left(\frac{h-h_{\rm c}}{B}\right)}
             {\left(\exp \left(\frac{h-h_{\rm c}}{B}\right)+1\right)^2},
\end{equation}

\noindent which has the form of a line shape function. One may
approximate the above with the pseudo-Epstein function:
\begin{equation} \label{PseudoEpsteinFunction}
  \begin{aligned}
    \tilde Ep(h,h_{\rm c},B) := &\frac{1}{16}
    \biggl\{
      \left[1+\left(\tfrac{h-h_{\rm c}}{2 B}\right)^2\right]^{-1}
      +\left[1+\left(\tfrac{h-h_{\rm c}}{4 B}\right)^2\right]^{-2}
      \\
      &
      +\left[1+\left(\tfrac{h-h_{\rm c}}{6 B}\right)^2\right]^{-3}
      +\left[1+\left(\tfrac{h-h_{\rm c}}{7 B}\right)^2\right]^{-4}
      \biggr\}^2,
  \end{aligned}
\end{equation}

\noindent which differs from the Epstein function by roughly one part in
$10^{3}$; this suffices, since the approximation for the
Appleton–Hartree equation is only valid to $3 \times 10^{-3}$ for GNSS
signal frequencies.

For simplicity, we construct a simple model for the electron density
$N_{\rm e}$ consisting of a sum of pseudo-Epstein functions:
\begin{equation} \label{IonosphericModel}
  \begin{aligned}
    N_{\rm e} :=& \biggl[
          \alpha_{\rm D} \, \tilde Ep(h,h_{\rm D},b_{\rm D})
          + \alpha_{\rm E} \, \tilde Ep(h,h_{\rm E},b_{\rm E}) \\
          & \>\>  + \alpha_{\rm F} \, \tilde Ep(h,h_{\rm F},b_{\rm F})\biggr].
  \end{aligned}
\end{equation}
The subscripts ${\rm D}$, ${\rm E}$, ${\rm F}$ on the parameters
correspond to the respective ionospheric layers. Of course the precise
profiles for the ionospheric layers are rather complicated and time
dependent, depending on the time of day and calendar date; in practice,
such detailed profiles are provided by the aforementioned ionospheric
models (see for instance \cite{EU2016ion} for a detailed description of
the NeQuick-G model employed in the Galileo system). However, a
simplified model for the ionospheric layers will suffice for
demonstrating the viability of our algorithm. We choose the parameter
values:
\begin{equation} \label{IndexRefractionCoeffs2}
  \begin{aligned}
    \alpha_{\rm D} &= 10^{12}~{\rm m}^{-3},
                   &\quad   h_{\rm D} &=  75~{\rm km},
                   &\quad   b_{\rm D} &=   5~{\rm km}, \\
    \alpha_{\rm E} &= 2.5 \times 10^{11}~{\rm m}^{-3},
                   &\quad   h_{\rm E} &= 130~{\rm km},
                   &\quad   b_{\rm E} &=  30~{\rm km}, \\
    \alpha_{\rm F} &= 10^{11}~{\rm m}^{-3},
                   &\quad   h_{\rm F} &= 300~{\rm km},
                   &\quad   b_{\rm F} &=  50~{\rm km} .
  \end{aligned}
\end{equation}

%
%
\subsection{Perturbation model} \label{sec:VC} To evaluate the potential
accuracy of our algorithm, we will introduce perturbations to simulate
the effect of uncertainties and errors in modeling the effective index
of refraction. The perturbations we introduce are simple rescalings of
the form:
\begin{equation} \label{PerturbationRescalings}
  \begin{aligned}
    n_{\rm eff} &= 1 + \Delta n_{\rm atm} \left(1
                 + \delta_1 \, \tilde p_1(h) \right) \\
    & \qquad 
    + \Delta n_{\rm ion} \left(1 + \delta_2 \, \tilde  p_2(h)\right),
  \end{aligned}
\end{equation}
where $\tilde p_A(h)$ ($A\in\{1,2\}$) denotes a perturbation function
$-1<\tilde p_A(h) <1$ and $\delta_1$ and $\delta_2$ correspond to the
respective fractional perturbations to $\Delta n_{\rm atm}$ and $\Delta
n_{\rm ion}$. For the tests, we choose $\tilde p_A(h)$ to have the form
\begin{equation} \label{PerturbationFunction}
  \begin{aligned}
    \tilde p_A(h) = \sum_{i} \alpha_i Ls(h,h_{0,i},\sigma_i),
  \end{aligned}
\end{equation}

\noindent where $\alpha_i$ are coefficients and $Ls(h,h_{0},\sigma)$
represents a line shape function centered at $h=h_{0}$ with a width
$\sigma$. We choose for the line shape function the following:
\begin{equation} \label{LineShape}
  \begin{aligned}
    Ls(h,h_{0},\sigma) = \frac{\sigma^2}{\sigma^2 
                      + (h-h_0)^2} \frac{\sigma^4}{\sigma^4 + (h-h_0)^4}
  \end{aligned}
\end{equation}

\noindent which qualitatively resembles a Lorentzian function, but with
a faster falloff. For the coefficients, we choose $\alpha_i=+1$ (to
maximize the refraction of the geodesics), and for $\vec h_0$ and
$\vec{\sigma}$, we choose for $p_1(h)$ the following:
\begin{equation} \label{PerturbationParametersAtm}
  \begin{aligned}
    \vec h_0        &= ( 0 , 4 , 8 , 12 , 16 ) ,\\
    \vec{\sigma}    &= ( 2.0 , 1.5 , 1.8 , 1.7 , 1.5 ),
  \end{aligned}
\end{equation}

\noindent and for $p_2(h)$
\begin{equation} \label{PerturbationParametersIon}
  \begin{aligned}
    \vec h_0        &= ( 150 , 200 , 250 , 300 , 350 ) ,\\
    \vec{\sigma}    &= ( 21 , 15 , 18 , 21 , 10 ) ,
  \end{aligned}
\end{equation}

\noindent all in units of ${\rm km}$.

We now discuss estimates for $\delta_1$, which corresponds to the
magnitude of fractional uncertainties in $\Delta n_{\rm atm}$ due to
variations in humidity and measurement uncertainties in the temperature
and pressure profiles near the surface of the Earth. Humidity variations
contribute $\sim 2 \times 10^{-4}$ in $\delta_1$ (see
\cite{SmithWeintraub1953} for a formula from which one may derive this
estimate). Achievable uncertainties \cite{WMO2018} of up to $\delta P =
15~{\rm Pa}$ and $\delta T = 0.2~{^\circ}{\rm C}$ at sea level
correspond to a contribution of $\sim 7 \times 10^{-4}$ in $\delta_0$.
After including humidity variations (with addition in quadrature), one
arrives at an uncertainty of $\sim 7.4 \times 10^{-4}$, which we round
up to obtain $\delta_1=10^{-3}$.

For uncertainties in the ionosphere, we consider several values for
$\delta_2$, the magnitude of fractional uncertainties in $\Delta n_{\rm
ion}$, which is determined by uncertainties in the ionospheric electron
density. One might expect to model the ionospheric electron density to
an accuracy of a few percent, as uncertainties of $<10\%$ in the total
electron content (the electron density integrated along a path) can in
principle be achieved \cite{Rovira2016}. In our tests, we will consider
values up to $\delta_2=0.10$, which corresponds to an uncertainty of
$10\%$ in the magnitude of fractional uncertainties in $\Delta n_{\rm
ion}$.


%
%

%
%
\section{Code test and benchmarks}\label{sec:VI}

%
%
\subsection{Test description}
The tests of the \textsc{squirrel.jl} code are performed in a manner
similar to the comparison tests in Sec. \ref{sec:MetricsKS} of the
\textsc{cereal.jl} code with Kerr-Schild geodesics. In particular, we
generated a set of $10^5$ target points $X_{\rm tar}$ on the WGS-84
reference ellipsoid and initial data for a spray of $N \geq 5$ null
geodesics from each of the target points (with the exception of the
benchmark tests, which include $N=4$ emission points). We consider up to
six emission points in our tests since GNSS satellite constellations are
typically designed with the requirement that six satellites are in view
at any given time \cite{Kaplan2017,Parkinson1996}. We then integrate
each geodesic to a radial coordinate value of $\sim 26.5 \times 10^3
{\rm km}$, the endpoints of which are then used as inputs for the
locator functions in the \textsc{squirrel.jl} code. The output of the
\textsc{squirrel.jl} code is then compared with the target points
$X_{\rm tar}$.

The effective geometry is described by the Gordon metric
\eqref{GordonMetric} with the effective index of refraction given by
Eqs. \eqref{EffectiveIndexofRefraction}, \eqref{IndexRefractionForm},
\eqref{IndexofRefractionIonospherePhase}, and
\eqref{IndexofRefractionIonosphere}. For the ``background'' spacetime
geometry, we use the weak field metric \eqref{WeakFieldMetric}, which
incorporates gravitational effects.

All test calculations were performed with double floating point
precision (\textsc{Float64}) to reduce execution time, though
\textsc{squirrel.jl} is written to accommodate extended precision
calculations (\textsc{DoubleFloats} \cite{DoubleFloats2018}, for
instance). For the generation of test cases, the tolerance (both
relative and absolute) for the ODE solvers is chosen to be $10^{-14}$,
and a high order integrator is employed, in particular the 9th order
\textsc{AutoVern9} in \textsc{OrdinaryDiffEq.jl} (as opposed to the the
7th order \textsc{AutoVern7} integrator used in the locator functions in
the \textsc{squirrel.jl} code). A basic validation test was performed
for the generation of test cases, where the outputs have been compared
for different tolerances; in all cases, relative differences were on the
order of machine precision. For the location code, the (user specified)
tolerances are chosen to be $10^{-10}$ to reduce execution time.

\begin{figure}
  \includegraphics[width=1.0\columnwidth]{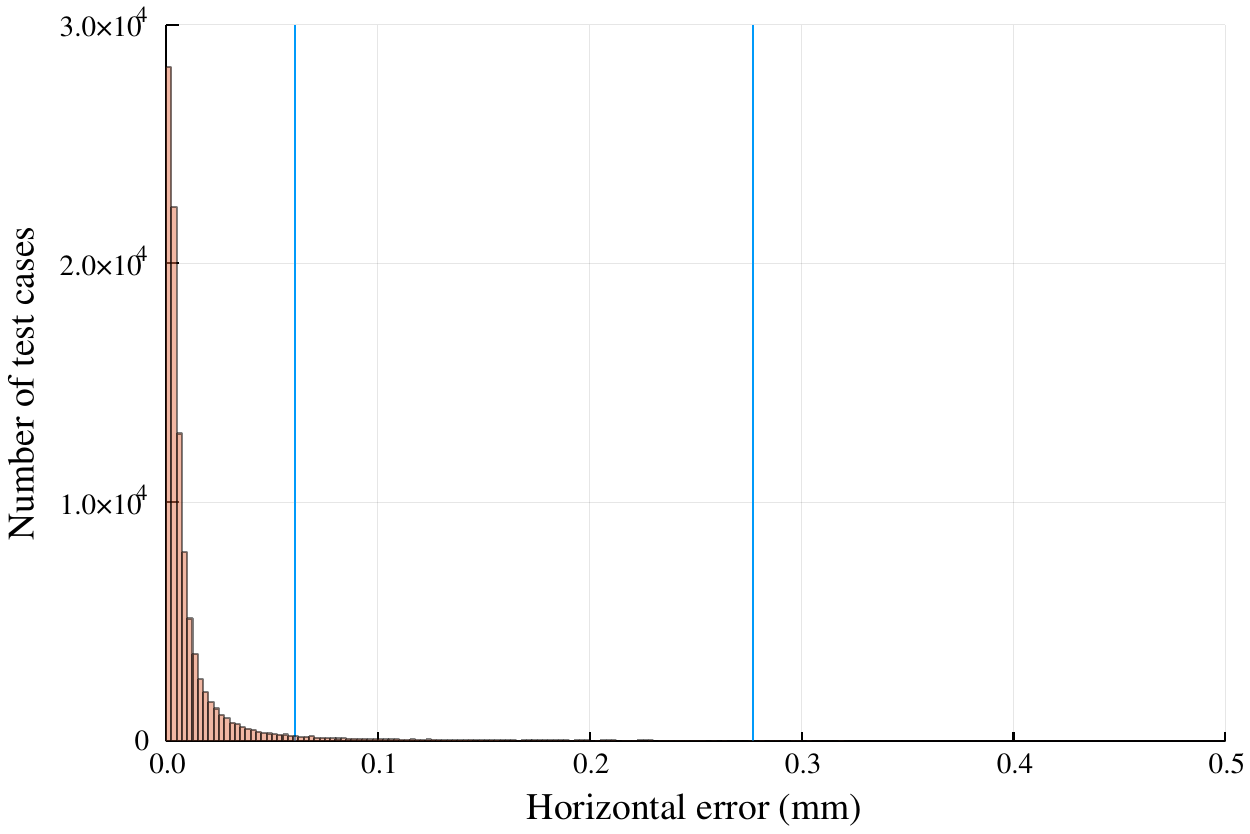}
  \caption{ Horizontal positioning errors ($n=5$ emission points) in the
    Kerr geometry for $10^5$ test cases. The vertical lines correspond
    to the $95\%$ confidence level and rms values for the errors, with
    respective values of $0.0608$ and $0.277~{\rm mm}$. Note that the
    results here are expressed in units of millimeters. Out of $10^5$
    samples, four samples ($0.004\%$) have an error $>2~{\rm cm}$ with
    the largest error being $3.39~{\rm cm}$. }
  \label{fig:plotHerr-n5k}
\end{figure}

\begin{figure}
  \includegraphics[width=1.0\columnwidth]{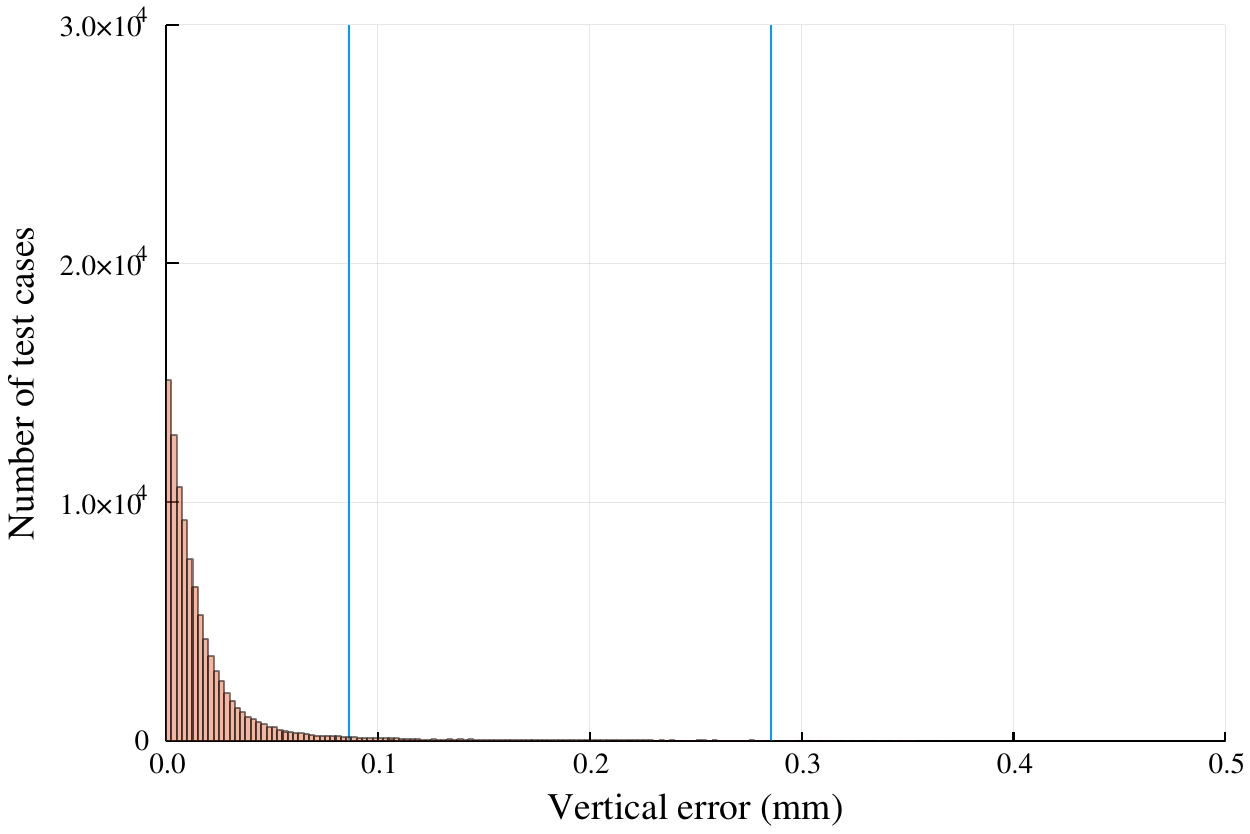}
  \caption{ Vertical positioning errors ($n=5$ emission points) in the
    Kerr geometry for $10^5$ test cases. The vertical lines correspond
    to the $95\%$ confidence level and rms values for the errors, with
    respective values of $0.0862$ and $0.286~{\rm mm}$. Note that the
    results here are expressed in units of millimeters. Out of $10^5$
    samples, $1$ sample ($0.001\%$) has an error $>2~{\rm cm}$ with the
    largest error being $2.40~{\rm cm}$. }
  \label{fig:plotVerr-n5k}
\end{figure}

\begin{figure}
  \includegraphics[width=1.0\columnwidth]{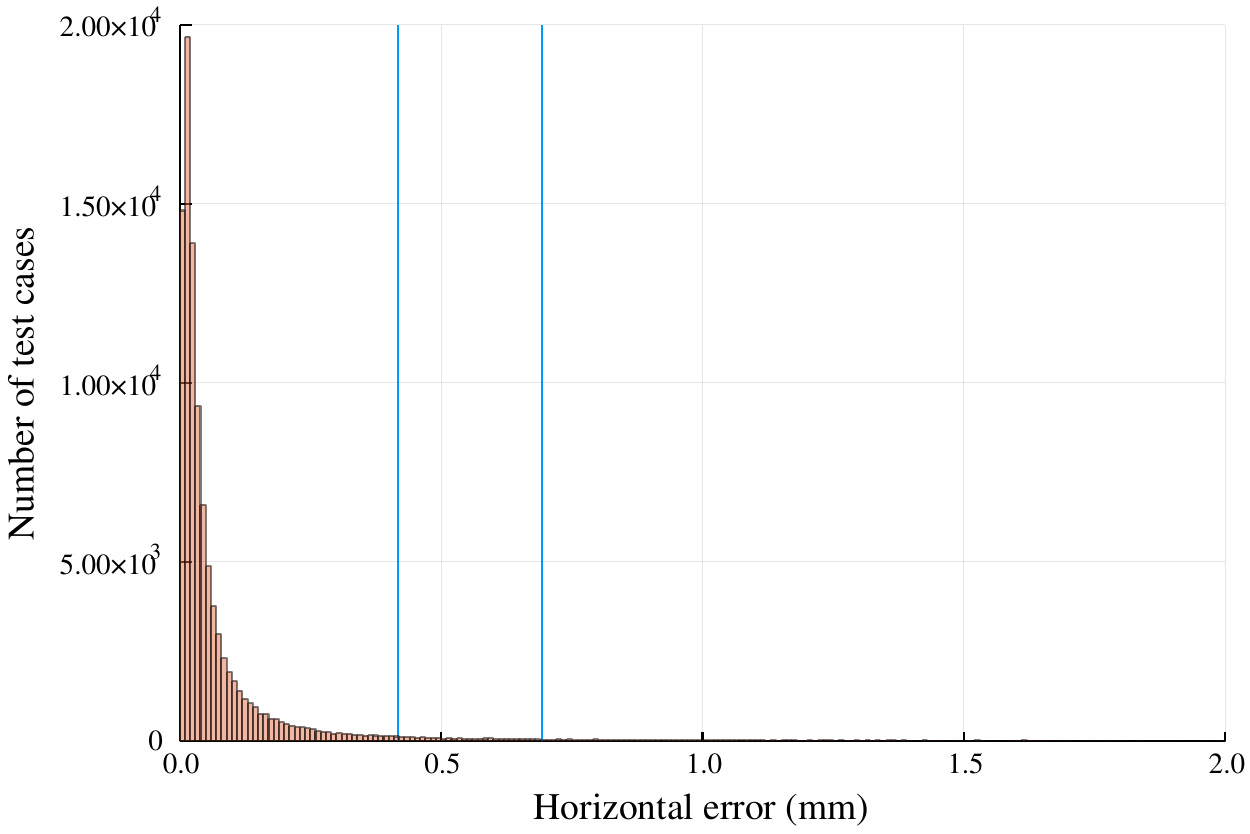}
  \caption{ Horizontal positioning errors ($n=5$ emission points) in the
    analog geometry incorporating tropospheric and ionospheric effects
    ($10^5$ test cases). The vertical lines correspond to the $95\%$
    confidence level and rms values for the errors, with respective
    values of $0.418$ and $0.693~{\rm mm}$. Out of $10^5$ samples, $13$
    samples ($0.013\%$) have an error $>2~{\rm cm}$ with the largest
    error being $3.20~{\rm cm}$. }
  \label{fig:plotHerr-n5p0}
\end{figure}

\begin{figure}
  \includegraphics[width=1.0\columnwidth]{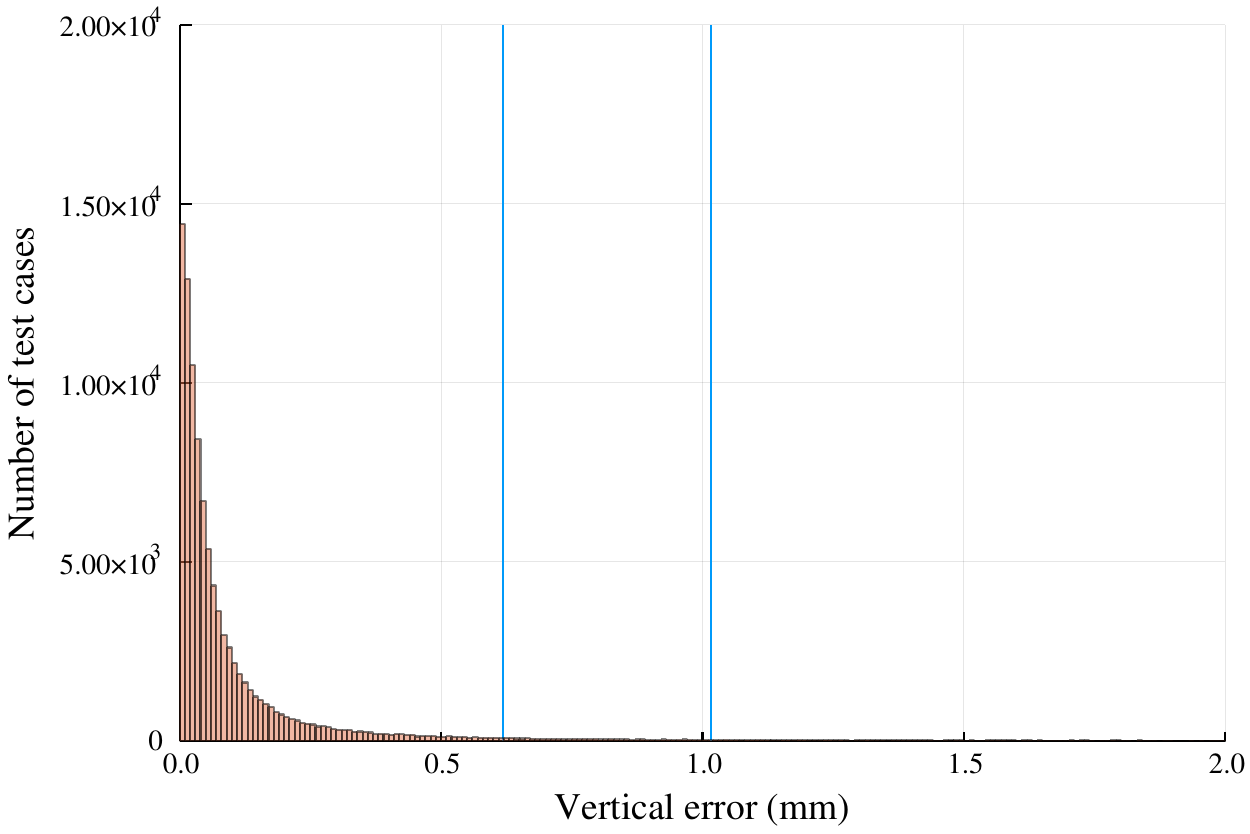}
  \caption{ Vertical positioning errors ($n=5$ emission points) in the
    analog geometry incorporating tropospheric and ionospheric effects
    ($10^5$ test cases). The vertical lines correspond to the $95\%$
    confidence level and rms values for the errors, with respective
    values of $0.618$ and $1.02~{\rm mm}$. Out of $10^5$ samples, $58$
    samples ($0.058\%$) have an error $>2~{\rm cm}$ with the largest
    error being $3.59~{\rm cm}$. }
  \label{fig:plotVerr-n5p0}
\end{figure}

\begin{figure}
  \includegraphics[width=1.0\columnwidth]{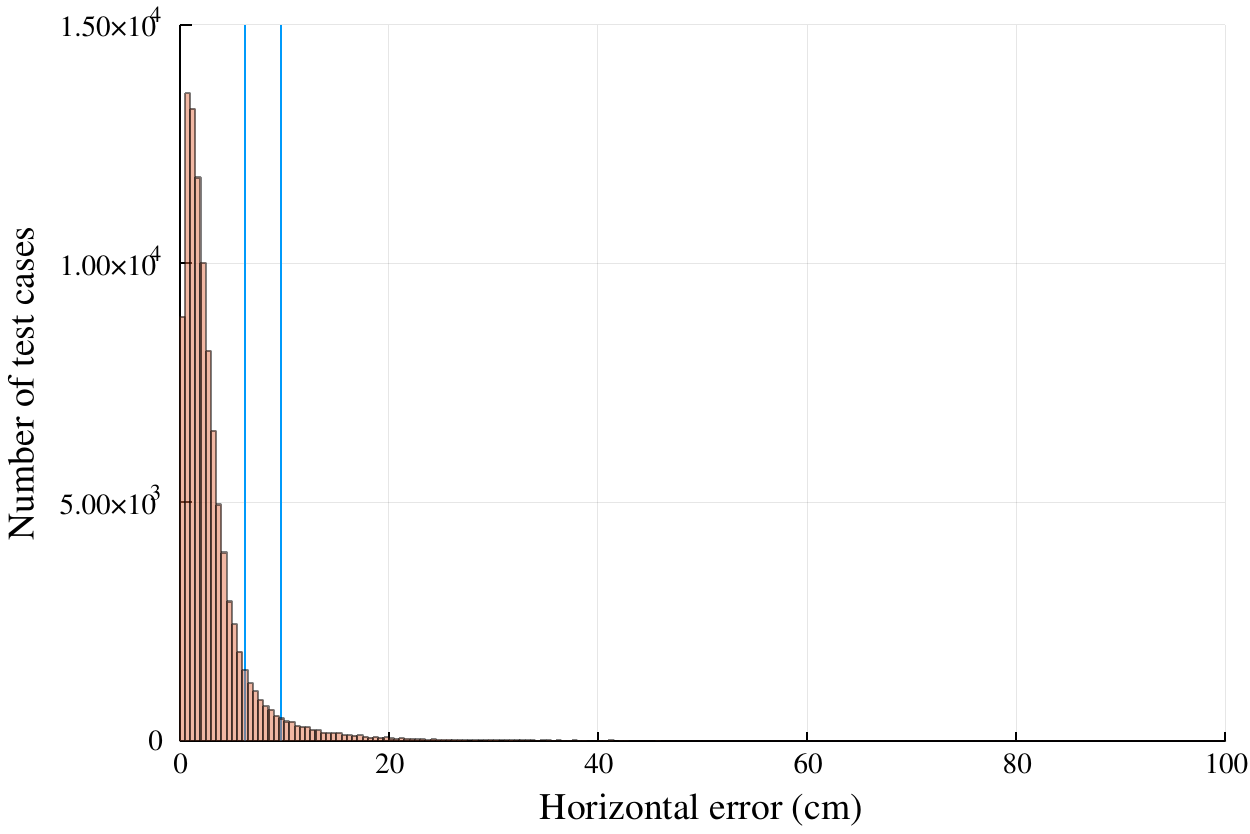}
  \caption{ Horizontal positioning errors ($n=5$ emission points) in the
    perturbed analog geometry corresponding to a fractional uncertainty
    of $0.1\%$ ($\delta_1=10^{-3}$) in the tropospheric index of
    refraction and $1\%$ ($\delta_2=0.01$) in the ionospheric electron
    density ($10^5$ test cases). The vertical lines correspond to the
    rms and $95\%$ confidence level values for the errors, with
    respective values of $6.27$ and $9.70~{\rm cm}$. Out of $10^5$
    samples, seven samples ($0.007\%$) have an error $>2~{\rm m}$ with
    the largest error being $3.42~{\rm m}$. }
  \label{fig:plotHerr-n5p1}
\end{figure}

\begin{figure}
  \includegraphics[width=1.0\columnwidth]{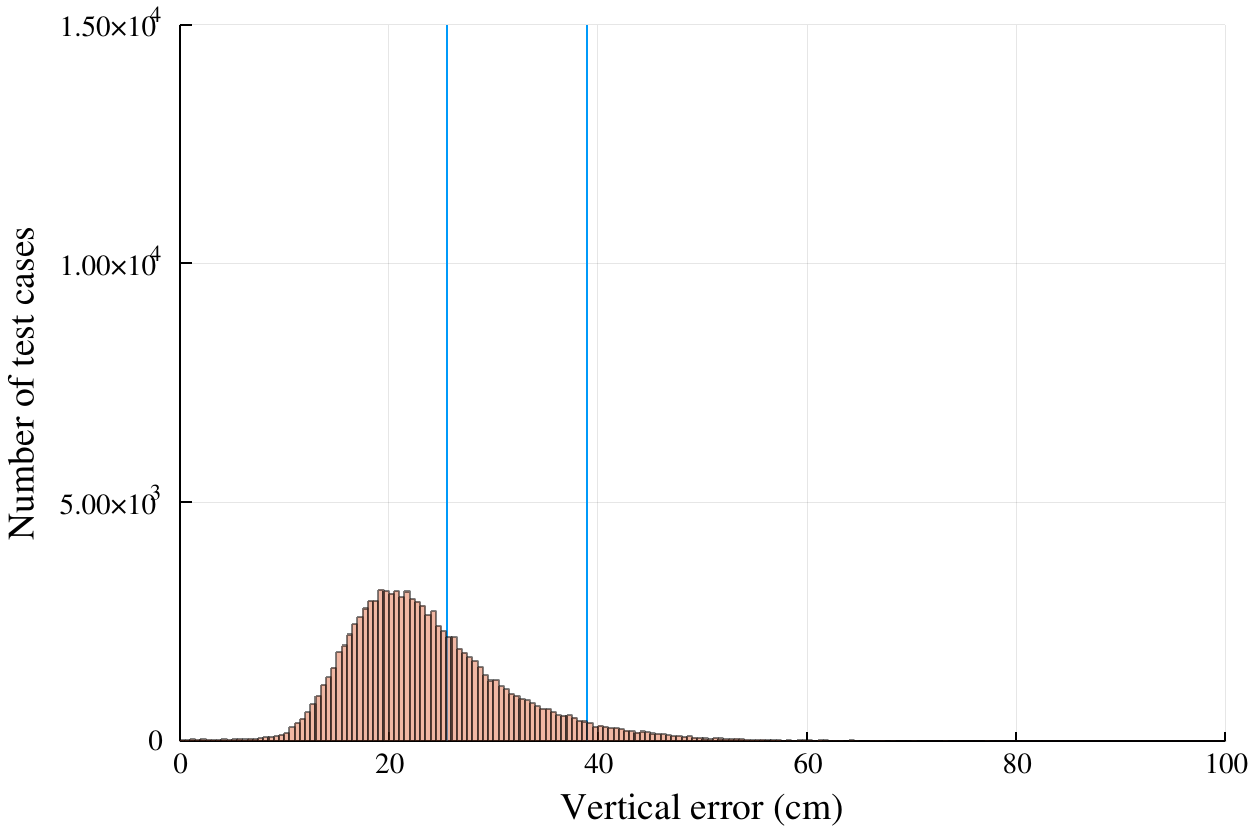}
  \caption{ Vertical positioning errors ($n=5$ emission points) in the
    perturbed analog geometry corresponding to a fractional uncertainty
    of $0.1\%$ ($\delta_1=10^{-3}$) in the tropospheric index of
    refraction and $1\%$ ($\delta_2=0.01$) in the ionospheric electron
    density ($10^5$ test cases). The vertical lines correspond to the
    RMS and $95\%$ confidence level values for the errors, with
    respective values of $25.6$ and $39.0~{\rm cm}$. Out of $10^5$
    samples, $10$ samples ($0.01\%$) have an error $>2~{\rm m}$ with the
    largest error being $4.56~{\rm m}$. }
  \label{fig:plotVerr-n5p1}
\end{figure}

\begin{figure}
  \includegraphics[width=1.0\columnwidth]{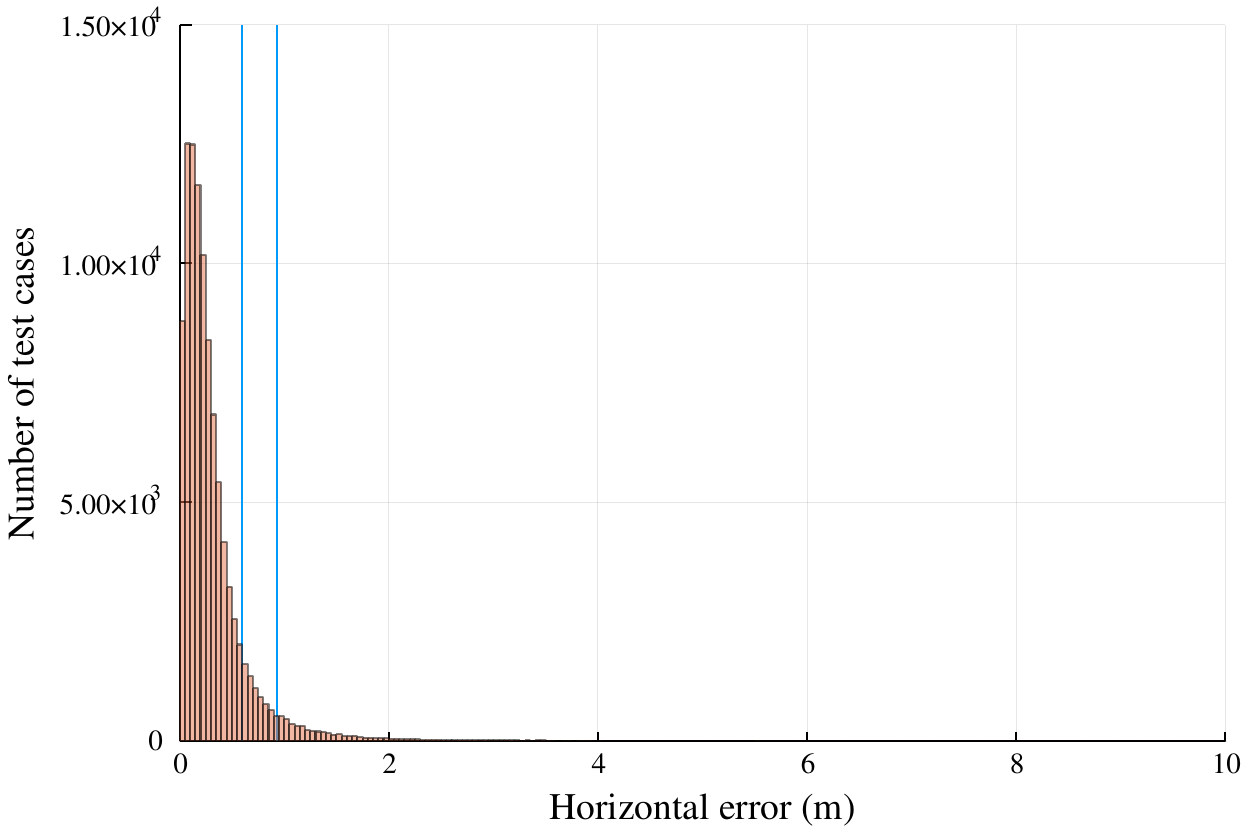}
  \caption{ Horizontal positioning errors ($n=5$ emission points) in the
    perturbed analog geometry corresponding to a fractional uncertainty
    of $0.1\%$ ($\delta_1=10^{-3}$) in the tropospheric index of
    refraction and $10\%$ ($\delta_2=0.1$) in the ionospheric electron
    density ($10^5$ test cases). The vertical lines correspond to the
    rms and $95\%$ confidence level values for the errors, with
    respective values of $59.4$ and $93.1~{\rm cm}$. Out of $10^5$
    samples, six samples ($0.006\%$) have an error $>20~{\rm m}$ with
    the largest error being $31.9~{\rm m}$. }
  \label{fig:plotHerr-n5p10}
\end{figure}

\begin{figure}
  \includegraphics[width=1.0\columnwidth]{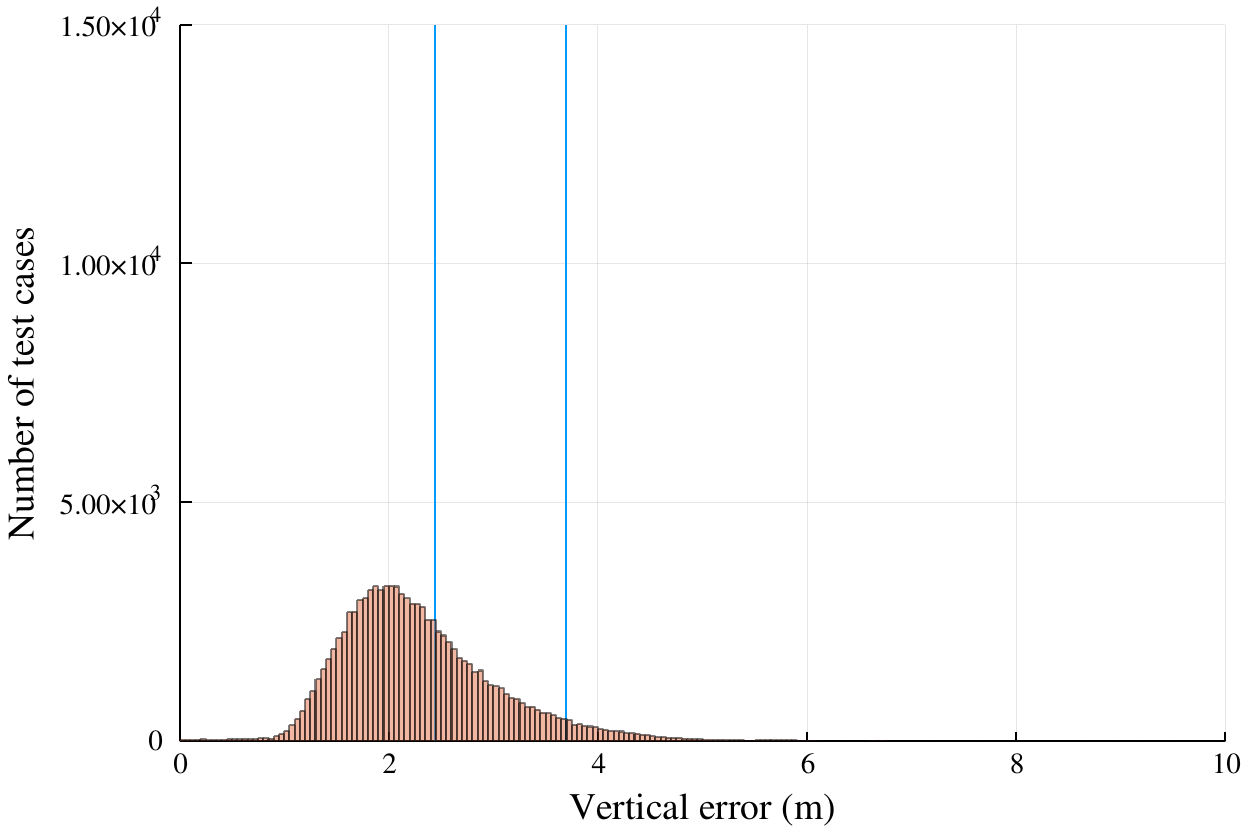}
  \caption{ Vertical positioning errors ($n=5$ emission points) in the
    perturbed analog geometry corresponding to a fractional uncertainty
    of $0.1\%$ ($\delta_1=10^{-3}$) in the tropospheric index of
    refraction and $10\%$ ($\delta_2=0.1$) in the ionospheric electron
    density ($10^5$ test cases). The vertical lines correspond to the
    rms and $95\%$ confidence level values for the errors, with
    respective values of $2.44$ and $3.70~{\rm m}$. Out of $10^5$
    samples, $10$ samples ($0.01\%$) have an error $>20~{\rm m}$ with
    the largest error being $42.5~{\rm m}$. }
  \label{fig:plotVerr-n5p10}
\end{figure}

%
%
\subsection{Test results}
Test results for $n=5$ emission points are presented in Figs.
\ref{fig:plotHerr-n5k}--\ref{fig:plotVerr-n5p10}. The vertical errors
correspond to errors projected in the direction orthogonal to the WGS-84
reference ellipsoid, and the horizontal errors correspond to errors
projected along directions tangent to the WGS-84 ellipsoid. For the Kerr
metric, the \textsc{squirrel.jl} code can achieve in most cases
submillimeter accuracy, demonstrating a potential for extreme precision
in vacuum environments; the methods presented in the
\textsc{squirrel.jl} code may be ideal for relativistic location in
space navigation. When tropospheric and ionospheric effects are
included, the \textsc{squirrel.jl} code yields errors on the order of a
centimeter, as illustrated in Figs. \ref{fig:plotHerr-n5p0} and
\ref{fig:plotVerr-n5p0}.

Figures \ref{fig:plotHerr-n5p1}--\ref{fig:plotVerr-n5p10} illustrate
positioning errors when including uncertainties in the determination of
the tropospheric and ionospheric index of refraction. These tests were
performed using the perturbation model described in the preceding
section; the test cases were generated with the unperturbed metric, and
for the tests themselves, the perturbed metric is used in the locator
functions of the \textsc{squirrel.jl} code (which take the metric
functions as an input). Errors resulting from an uncertainty of $1\%$ in
the determination of the ionospheric refractive index are illustrated in
Figs. \ref{fig:plotHerr-n5p1} and \ref{fig:plotVerr-n5p1}; this is the
best result one can realistically expect to achieve with the
approximation \eqref{IndexofRefractionIonospherePhase} for GNSS signal
frequencies of $\sim 1~{\rm GHz}$ (but we reiterate that higher signal
frequencies can achieve improved accuracy with the same approximation).
Even then, the horizontal positioning errors are for the most part
confined to less than $\sim 10~{\rm cm}$ to a $95\%$ confidence level,
while the vertical errors exhibit a systematic shift of $\sim 20~{\rm
cm}$, which corresponds to the fact that the perturbations to the index
of refraction in our perturbation model are positive, and that the index
of refraction profiles vary primarily in the radial direction. 

Errors from an uncertainty of $10\%$ in the ionospheric profile are
illustrated in Figs. \ref{fig:plotHerr-n5p10} and
\ref{fig:plotVerr-n5p10}, which increases the horizontal errors to $\sim
1~{\rm m}$ and the systematic shift in the vertical errors to $\sim
2~{\rm m}$. Upon comparison with the single frequency errors reported in
the latest Galileo quarterly report \cite{EU2021} for the first three
months of 2021, we note that the rms and $95\%$ confidence level values
are somewhat comparable to the performance of Galileo (rms and $95\%$
C.L. $\sim 1 \, {\rm m}$), albeit for a smaller sample size in our case.
Of course, the results presented here do not take into account other
GNSS errors, such as multipath, satellite timing and ephemeris errors,
the latter two of which have been addressed in
\cite{Colmenero:2021fbq,Kostic:2015cca,Gomboc2014,Cadez2010,Delva:2010zx}.
In the Galileo error budget, such errors [referred to as signal in space
errors (SISE)] are on the order of half a meter \cite{EU2021}, and, for
single-frequency users, are smaller in magnitude than the error
contributions from ionospheric and tropospheric effects. To compare the
errors presented in this article with those of \cite{EU2021}, one should
add in quadrature a $\sim 0.5 {\rm m}$ contribution from SISE; even with
such a correction to the rms and $95\%$ C.L. values presented in Fig.
\ref{fig:plotHerr-n5p10} for horizontal positioning, our corrected
errors ($0.776~{\rm m}~[rms]$ and $1.06~{\rm m}~[95\% C.L.]$) remain
smaller than those reported in \cite{EU2021}, and with fewer errors
$\geq 20~{\rm m}$ in proportion.  This indicates a potential for
improved performance, even with an uncertainty\footnote{An interesting
question worth investigating (left for future work) is whether one can
construct simple, high-accuracy ionospheric models which reduce this
uncertainty---see \cite{Rovira2016,Rovira2016fast}.} of $10\%$ in the
determination of the ionospheric free electron density and for the
stated uncertainties in the determination of atmospheric parameters in
the lower troposphere. Moreover, reduced uncertainties in the
determination of the ionospheric electron density profile to the $1\%$
level can reduce the $95\%$ C.L. errors by a factor of $10$, to roughly
a decimeter.

\begin{figure}
  \includegraphics[width=1.0\columnwidth]{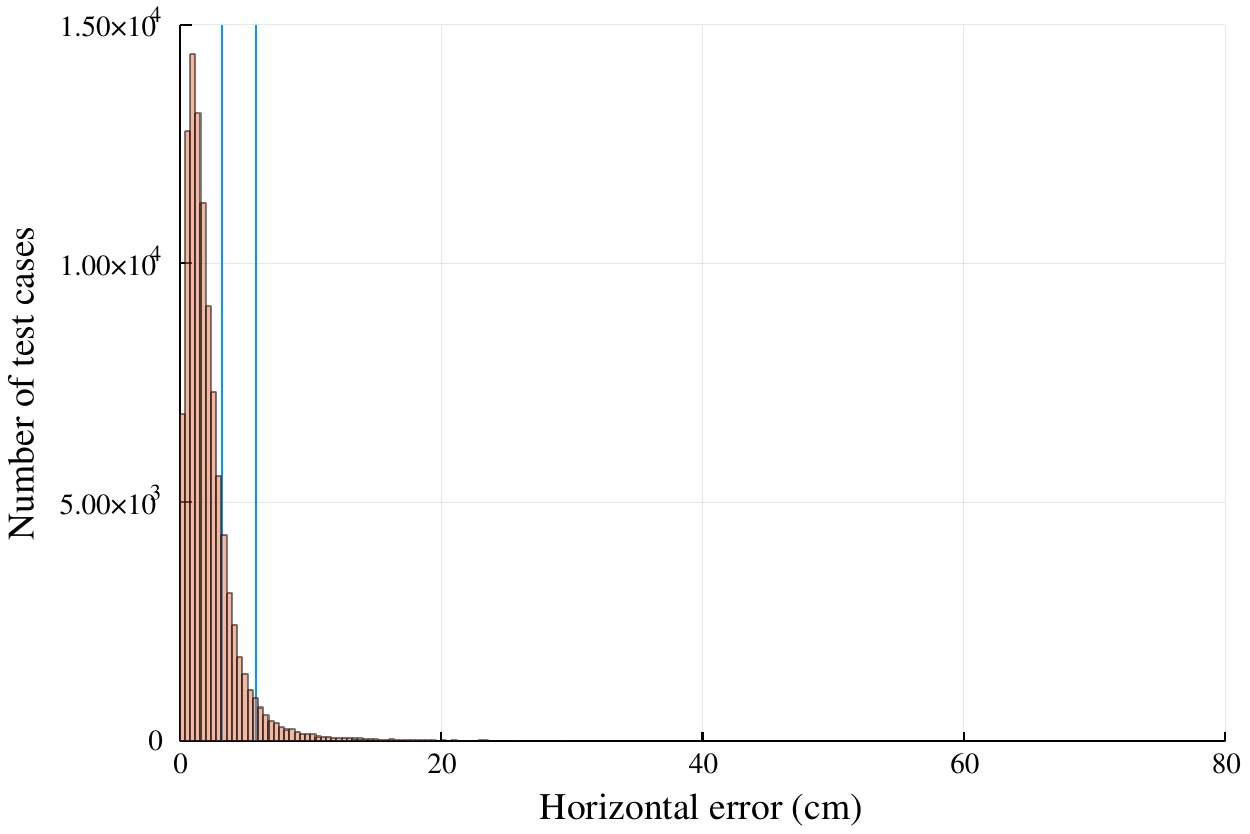}
  \caption{ Horizontal positioning errors ($n=6$ emission points) in the
    perturbed analog geometry corresponding to a fractional uncertainty
    of $0.1\%$ ($\delta_1=10^{-3}$) in the tropospheric index of
    refraction and $1\%$ ($\delta_2=0.01$) in the ionospheric electron
    density ($10^5$ test cases). The vertical lines correspond to the
    rms and $95\%$ confidence level values for the errors, with
    respective values of $3.27$ and $5.81~{\rm cm}$. Out of $10^5$
    samples, $1$ sample ($0.001\%$) has an error $>1~{\rm m}$ (none
    greater than $2~{\rm m}$) with the largest error being $1.56~{\rm
    m}$. }
  \label{fig:plotHerr-n6p1}
\end{figure}

\begin{figure}
  \includegraphics[width=1.0\columnwidth]{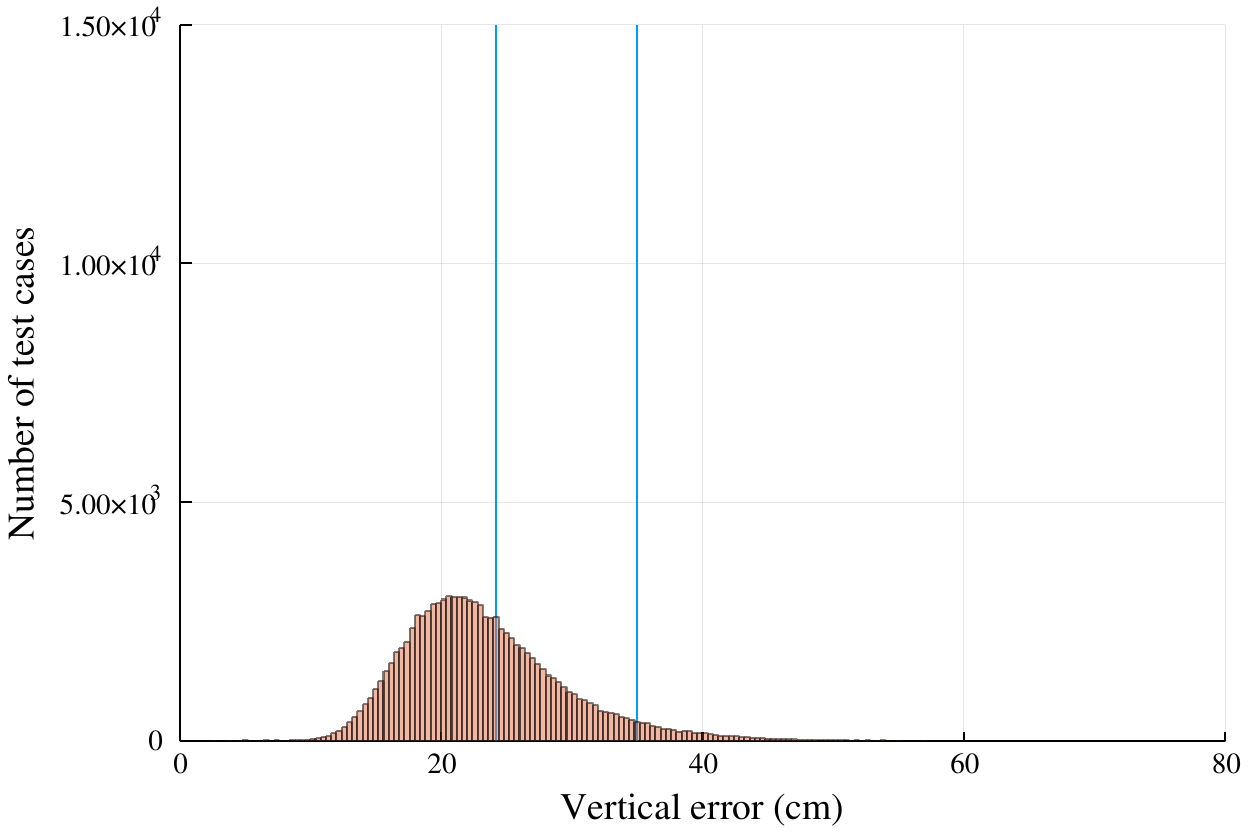}
  \caption{ Vertical positioning errors ($n=6$ emission points) in the
    perturbed analog geometry corresponding to a fractional uncertainty
    of $0.1\%$ ($\delta_1=10^{-3}$) in the tropospheric index of
    refraction and $1\%$ ($\delta_2=0.01$) in the ionospheric electron
    density ($10^5$ test cases). The vertical lines correspond to the
    rms and $95\%$ confidence level values for the errors, with
    respective values of $24.2$ and $35.0~{\rm cm}$. Out of $10^5$
    samples, $1$ sample ($0.001\%$) have an error $>1~{\rm m}$ (none
    greater than $2~{\rm m}$) with the largest error being $1.10~{\rm
    m}$. }
  \label{fig:plotVerr-n6p1}
\end{figure}

\begin{figure}
  \includegraphics[width=1.0\columnwidth]{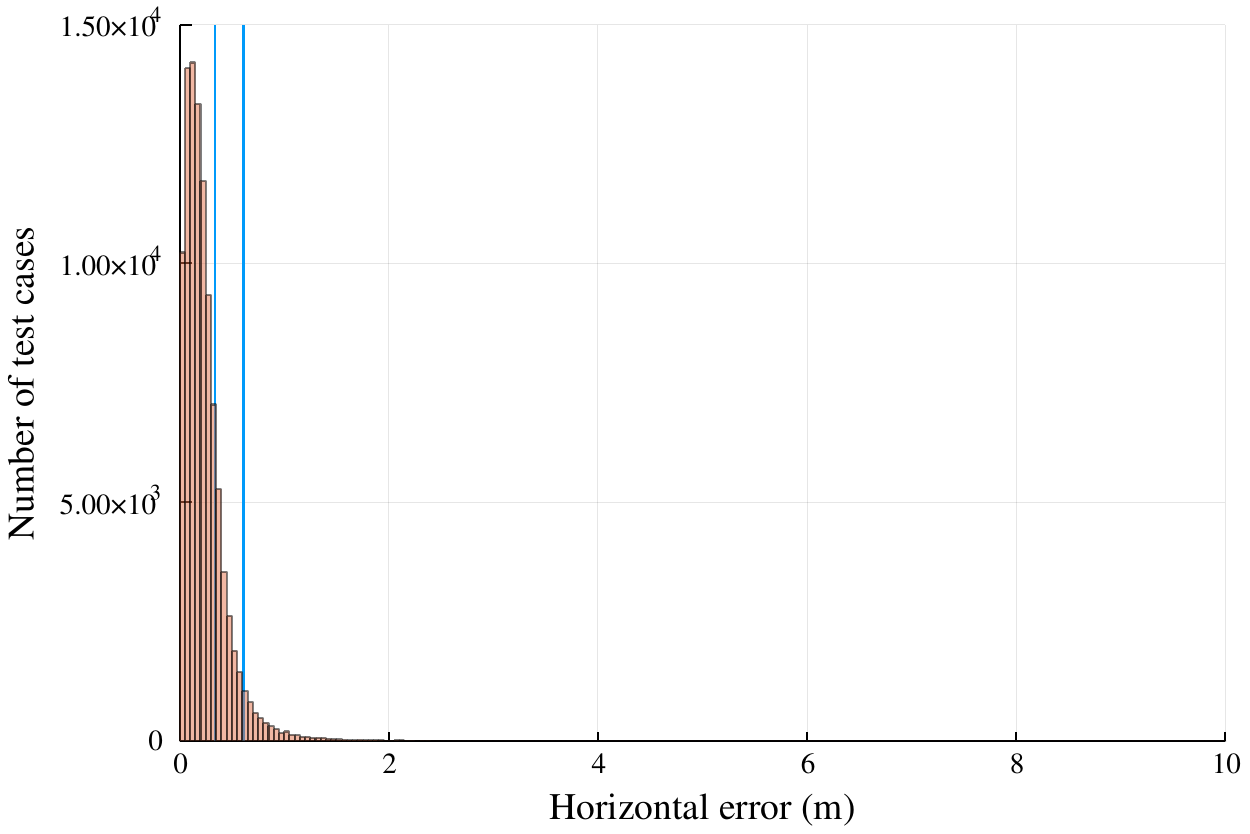}
  \caption{ Horizontal positioning errors ($n=6$ emission points) in the
    perturbed analog geometry corresponding to a fractional uncertainty
    of $0.1\%$ ($\delta_1=10^{-3}$) in the tropospheric index of
    refraction and $10\%$ ($\delta_2=0.1$) in the ionospheric electron
    density ($10^5$ test cases). The vertical lines correspond to the
    rms and $95\%$ confidence level values for the errors, with
    respective values of $33.5$ and $61.1~{\rm cm}$. Out of $10^5$
    samples, seven samples ($0.007\%$) samples have an error $>5~{\rm
    m}$ with the largest error being $14.6~{\rm m}$. }
  \label{fig:plotHerr-n6p10}
\end{figure}

\begin{figure}
  \includegraphics[width=1.0\columnwidth]{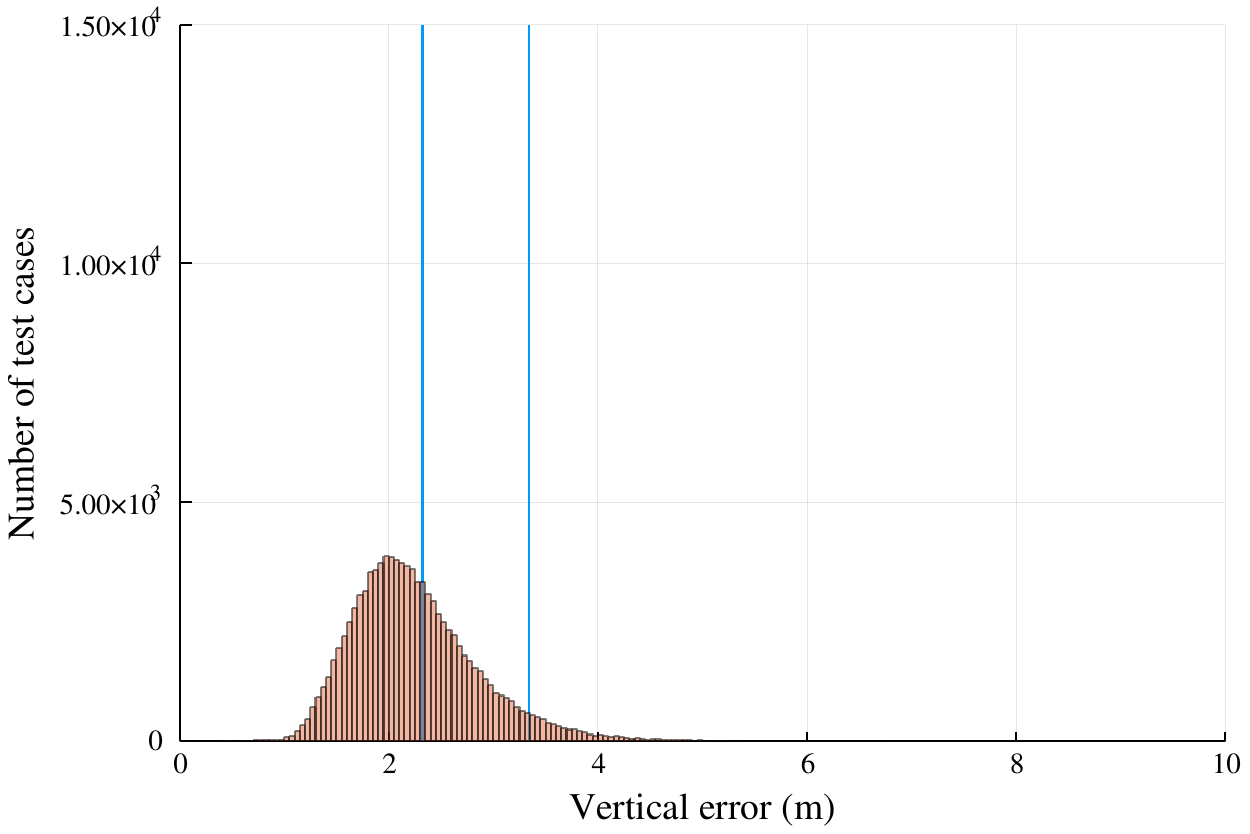}
  \caption{ Vertical positioning errors ($n=6$ emission points) in the
    perturbed analog geometry corresponding to a fractional uncertainty
    of $0.1\%$ ($\delta_1=10^{-3}$) in the tropospheric index of
    refraction and $10\%$ ($\delta_2=0.1$) in the ionospheric electron
    density ($10^5$ test cases). The vertical lines correspond to the
    rms and $95\%$ confidence level values for the errors, with
    respective values of $2.32$ and $3.24~{\rm m}$. Out of $10^5$
    samples, five samples ($0.005\%$) samples have an error $>7~{\rm m}$
    with the largest error being $10.4~{\rm m}$. }
  \label{fig:plotVerr-n6p10}
\end{figure}

One can obtain improved accuracy with additional emission points. With
$n=6$ emission points, the largest errors are significantly reduced, as
indicated in the figures \ref{fig:plotHerr-n6p1}--
\ref{fig:plotVerr-n6p10} for the cases with $1\%$ and $10\%$ fractional
uncertainty in the ionospheric electron density $N_{\rm e}$. These two
cases are chosen since they form the boundary cases for the accuracy
that one might expect to be achievable with the \textsc{squirrel.jl}
code at GNSS signal frequencies of $\sim 1~{\rm GHz}$. In both cases, we
find a significant reduction in the number of large errors for $n=6$
emission points. There were no errors above the stated thresholds for
the $n=5$ cases ($2~{\rm m}$ for the $1\%$ case and $20~{\rm m}$ for the
$10\%$ case); for the $n=6$ cases, we find only one sample with an error
$>1~{\rm m}$ threshold for a $1\%$ uncertainty in $N_{\rm e}$, and seven
samples with horizontal errors $>5~{\rm m}$ a $10\%$ uncertainty in
$N_{\rm e}$. The rms and $95\%$ confidence level errors are roughly the
same for the vertical errors (owing to the systematic shift that the
perturbations introduce), but are reduced by a factor of $2$ for the
horizontal errors. This result indicates that, in principle, the
inclusion of additional emission points can significantly reduce the
number of large errors in the \textsc{squirrel.jl} code.

%
%
\subsection{Benchmarks}\label{sec:Benchmarks}

\begin{table}
\begin{tabular}{ |p{1cm}||p{2.2cm}|p{2.2cm}|p{2.2cm}|  }
 \hline
 \multicolumn{4}{|c|}{Execution time for \textsc{squirrel.jl}} \\
 \hline
 $N$ & Kerr-Schild & Gordon & Pert. Gordon\\
 \hline
 4 & 27 ms  & 193 ms & 223 ms\\
 5 & 101 ms & 553 ms & 588 ms\\
 6 & 358 ms & 1.74 s & 1.95 s\\
 \hline
\end{tabular}
  \caption{Benchmarks performed with \textsc{squirrel.jl}. $N$ is the
  number of emission points. Tests were performed with the Kerr-Schild
  metric, the unperturbed Gordon metric, and the perturbed Gordon metric
  corresponding to a $10\%$ uncertainty in the ionospheric profile.}
\label{fig:Benchmark}
\end{table}

Some basic benchmarks have been performed for various situations; the
results are displayed in Table \ref{fig:Benchmark}. The results we
report were obtained on standard desktop computer with an Intel i5-7500
processor, and with four threads enabled. The benchmarks were performed
for three geometries: the Kerr-Schild metric, the unperturbed Gordon
metric (representing tropospheric and ionospheric effects), and the
perturbed Gordon metric corresponding to a $10\%$ uncertainty in the
ionospheric profile. We consider three cases, with $N=4,5,6$ emission
points. We note that the $N=4$ Kerr-Schild case has an execution time
comparable to that reported in \cite{Gomboc2014,Kostic:2015cca} for a
Schwarzschild location method. In the case of $N=4$ emission points, the
flat spacetime methods of \cite{Coll:2009tm,Coll:2012dk} and Sec.
\ref{sec:IIB} return two guesses due to the bifurcation problem, so the
squirrel algorithm is applied twice (one for each guess) for $N=4$
emission points. This is seen in the fact that the execution time for
$N=4$ is longer than one might expect from the number of combinations
$C(5,4)=5$, which is supported by the execution times for the Gordon and
perturbed Gordon cases.\footnote{The discrepancy in the Kerr-Schild case
may be due to overhead related to the different methods employed by the
\textsc{squirrel.jl} code between the $N=4$ and $N=5$ cases} Comparing
the $N=5$ and $N=6$ cases, we find a scaling roughly consistent with the
number of combinations $C(5,4)=5$, $C(6,4)=15$, which suggests an
increase in computational complexity by a factor of $3$.


%
%
%
%

\section{Summary and discussion}\label{sec:VII}

In this article, we have described and demonstrated a new method for
relativistic location in slightly curved, but otherwise generic
spacetime geometries. Though such methods may be of primary interest for
high precision space navigation in regions beyond the ionosphere, we
have demonstrated, by way of simple analog gravity models, that our
method can nonetheless be used to incorporate tropospheric and
ionospheric effects in terrestrial positioning. Though one might regard
such effects as ancillary from a purely general relativistic
perspective, we argue that they are still of fundamental importance in
the sense that the placement of emission coordinates near the surface of
the Earth will depend on the knowledge of the profile for the effective
tropospheric and ionospheric refractive index.

The methods we have described and implemented \cite{FHC_RPSsupp} make
use of state of the art automatic differentiation and ODE libraries
available in the Julia language, which permit the efficient evaluation
of the derivatives of numerical solutions of the geodesic equation
performed with respect to initial data. Combined with a quasi-Newton
root-finding algorithm, we have demonstrated that our methods can, with
guesses provided by the flat spacetime relativistic location formula of   
\cite{Ruggiero:2022}, accurately and efficiently compute the
intersection point of future pointing null cones from a set of spacelike
separated emission points. In particular, our implementation, the
\textsc{squirrel.jl} code \cite{FHC2022scurl}, can with five emission
points achieve submillimeter accuracy for terrestrial positioning
(satellite orbits at $\sim 26.5 \times 10^3 ~ {\rm km}$, target point at
surface of Earth) in a vacuum Kerr-Schild metric. When tropospheric and
ionospheric effects are included by way of the Gordon metric, the
\textsc{squirrel.jl} code can achieve horizontal errors of less than
$\sim 1~{\rm m}$ (according to rms and $95\%$ C.L. values for $10^5$
samples) for a $10\%$ uncertainty in the ionospheric free electron
density profile, and to less than $\sim 10~{\rm cm}$ for a $1\%$
uncertainty, with an execution time of $< 1~{\rm s}$ on a desktop
computer for five emission points. An interesting question for future
investigation is whether multifrequency methods may be used in
conjunction with our algorithm to constrain the electron density
profile.

Our test results indicate that the relativistic location algorithm
implemented in the \textsc{squirrel.jl} code can achieve extreme
precision for space navigation in vacuum regions beyond the ionosphere;
we will describe in detail the applications of our methods to deep space
navigation elsewhere. Our tests also indicate that implementations of
our method have the potential for performance comparable to or exceeding
the single-frequency performance of Galileo, assuming that the local
atmospheric properties of the troposphere are known to typical
measurement uncertainties and the ionospheric free electron profile is
known to an uncertainty of $10\%$ or less. Alternatively, the methods
presented in this article may perhaps be of interest as an additional
method for atmospheric tomography (see \cite{Jin2014gnss} and references
therein for an overview of methods in GNSS tomography).

The results we have presented here are complementary to those of
\cite{Colmenero:2021fbq,Kostic:2015cca,Gomboc2014,Cadez2010,Delva:2010zx},
which address the problem of incorporating general relativistic effects
in the determination of satellite ephemerides; satellite ephemeris and
timing errors (SISE) are some the largest contributors to GNSS error
budgets. A more comprehensive collection of methods for relativistic
positioning will require at the minimum a relativistic location
algorithm and an algorithm for determining satellite ephemerides and for
mitigating clock errors. These issues can in principle be addressed in a
more general optimization framework based on emission coordinates, as
discussed in \cite{Tarantola:2009hk}; the implementation of this
framework using modern machine learning methods will be explored in
future work.


%
%


\begin{acknowledgments}
We thank Miguel Zilh{\~a}o introducing us to the Julia language and his
invaluable guidance and advice. We also thank Taishi Ikeda, David
Hilditch, Edgar Gasper\'in, Chinmoy Bhattacharjee, and Richard Matzner
for helpful discussions and suggestions. J.C.F. also thanks DIME at the
University of Genoa for hosting a visit during which part of this work
was performed, and acknowledges financial support from FCT---Funda\c
c\~ao para a Ci\^encia e a Tecnologia of Portugal Grant
No.~PTDC/MAT-APL/30043/2017 and Project No.~UIDB/00099/2020. The work of
F.H. is supported by the Czech Science Foundation GA{\v C}R, Project No.
20-16531Y.
\end{acknowledgments}


\bibliography{squirrel}


\end{document}